\documentclass[preprint,12pt]{elsarticle}

\usepackage{amssymb}
\usepackage{amsmath}
\usepackage{lipsum}
\usepackage{mathptmx}
\usepackage{multirow}
\usepackage{tabularx}
\usepackage{hyperref}
\usepackage{array}   
\usepackage{diagbox} 
\usepackage{array}   
\usepackage{caption} 
\usepackage{booktabs}
\journal{Journal of Information Security and Applications}

\begin{document}

\begin{frontmatter}



\title{Examining Attacks on Consensus and Incentive Systems in Proof-of-Work Blockchains: A Systematic Literature Review}

\author[first]{Dinitha Wijewardhana}
\affiliation[first]{organization={University of Ruhuna },%Department and Organization
            addressline={A2}, 
            city={Wellamadama},
            state={Matara},
            postcode={81000}, 
            country={Sri Lanka}}
\author[first]{Sugandima Vidanagamachchi}
% \affiliation[first]{organization={University of Ruhuna },%Department and Organization
%             addressline={A2}, 
%             city={Wellamadama},
%             postcode={81000}, 
%             state={Matara},
%             country={Sri Lanka}}
\author[second]{Nalin Arachchilage}
\affiliation[second]{organization={ RMIT University },%Department and Organization
            addressline={124 La Trobe St}, 
            city={Melbourne}, 
            state={VIC},
            postcode={3000},
            country={ Australia }}
\begin{abstract}
Cryptocurrencies have gained popularity due to their transparency, security, and accessibility compared to traditional financial systems, with Bitcoin, introduced in 2009, leading the market. Bitcoin’s security relies on blockchain technology—a decentralized ledger consisting of a consensus and an incentive mechanism. The consensus mechanism, Proof of Work (PoW), requires miners to solve difficult cryptographic puzzles to add new blocks, while the incentive mechanism rewards them with newly minted bitcoins. However, as Bitcoin’s acceptance grows, it faces increasing threats from attacks targeting these mechanisms, such as selfish mining, double-spending, and block withholding. These attacks compromise security, efficiency, and reward distribution. Recent research shows that these attacks can be combined with each other or with either malicious strategies, such as network-layer attacks, or non-malicious strategies, like honest mining. These combinations lead to more sophisticated attacks, increasing the attacker’s success rates and profitability. Therefore, understanding and evaluating these attacks is essential for developing effective countermeasures and ensuring the long-term security. This paper begins by examining the  individual attacks executed in isolation and their profitability. It then explores how combining these attacks with each other or with other malicious and non-malicious strategies can enhance their overall effectiveness and profitability. The analysis further explores how the deployment of attacks such as selfish mining and block withholding by multiple competing mining pools against each other impacts their economic returns. Lastly, a set of design guidelines is provided, outlining areas future work should focus on to prevent or mitigate the identified threats.
\end{abstract}


% \begin{highlights}
% \end{highlights}


\end{frontmatter}

\section{Introduction}
\label{introduction}

Cryptocurrency is a type of digital or virtual money that runs on decentralized networks, which are not under the jurisdiction of a single centralized authority like a bank or government \citep{jafari2018cryptocurrency, aziz2022money, icta}. Currently, there are thousands of different cryptocurrencies available which can be used for investing, enabling smart contracts, powering decentralized applications, facilitating peer-to-peer transactions, and taking part in decentralized finance (DeFi) systems \citep{zheng2020overview,makarov2022cryptocurrencies,lee2018cryptocurrency}. The market leader and original cryptocurrency in the cryptocurrency market is Bitcoin \citep{nakamoto2008bitcoin} which makes up 48.6\% of the total value of the crypto market as of 2024. As of February 2024, the global cryptocurrency market cap is USD 2.09 trillion whereas Bitcoin’s market cap of USD 1.02 trillion accounts for around 50\% of that total \footnote{\href{https://www.coingecko.com/}{Cryptocurrency Prices, Charts, and Crypto Market Cap \citep{fang2021defi,glenski2019characterizing,vidal2022cryptocurrency}}}. During the global economic crisis of 2009, Bitcoin was introduced as a solution to the issues with centralized transaction management. It offers a number of advantages, including increased trust, security, and transparency among member organizations by enhancing the traceability of data shared across a business network and generating cost savings through new efficiencies \citep{antonopoulos2014mastering}. Nearly all cryptocurrencies, including Bitcoin \citep{nakamoto2008bitcoin}, Ethereum \citep{wood2014ethereum}, Bitcoin Cash \citep{cash2019bitcoin}, and Litecoin \citep{jumaili2021comparison}, are secured by blockchain networks. A blockchain is essentially a public ledger of transactions that anybody can examine and validate \citep{pilkington2016blockchain}. Transactions are broadcast by users in a peer-to-peer network, and participants use this ledger to validate them. The decentralization of the  blockchain among a network of nodes ensures that it is not under the control of a single entity \citep{tschorsch2016bitcoin}. 

The consensus and incentive mechanisms are two of the core components of blockchain networks \citep{tschorsch2016bitcoin,alyaseen2019consensus,wang2019survey}. As the blockchain is decentralized, a consensus mechanism is essential for achieving common agreement among all nodes on the state of the ledger, thereby preventing inconsistencies and fraudulent updates \citep{shrimali2022blockchain,bodkhe2020survey}. There are several consensus mechanisms, each with its own approach. These include Proof-of-Work (PoW) \citep{tschorsch2016bitcoin}, Proof-of-Stake (PoS) \citep{king2012ppcoin}, Proof-of-Activity (PoA) \citep{bentov2014proof}, and Proof-of-Burn (PoB) \citep{King2012PPCoinPC}, among others. Bitcoin employs PoW consensus mechanism which involves, participants, known as miners, compete to solve a complex cryptographic puzzle known as the PoW puzzle \citep{alyaseen2019consensus, mingxiao2017review}. When a miner successfully solves this puzzle, they share the solution with the network. The other nodes in the network then check if the solution is correct. After confirmation, the new block is appended to the blockchain. This process is referred to as mining \citep{nguyen2018survey}.

The incentive mechanism in blockchain is responsible for issuing and distributing rewards \citep{wen2021attacks}. Incentives are financial rewards provided by the system to motivate miners to participate in the mining process and verify transactions \citep{cao2019bitcoin, han2022can}. The issuing mechanism specifies how new cryptocurrency tokens are created. In many blockchain systems, miners who successfully validate transactions and add new blocks to the blockchain are rewarded with newly created units of the digital currency \citep{marchesi2022blockchain}. In the context of Bitcoin, there are two main sources of incentives: \textit{mining} and \textit{transaction fees} \citep{tschorsch2016bitcoin}. The miner who successfully solves the PoW puzzle receives a reward, which consists of newly issued Bitcoins \citep{chaudhry2018consensus,tschorsch2016bitcoin}. Interestingly, the term "mining" is used to describe this process because, much like digging for precious metals, it involves a resource-intensive effort to uncover valuable newly minted Bitcoins through complex computations \citep{tschorsch2016bitcoin}. The other source of incentives is the transaction fees for the transactions miners include in a block. These transaction fees are charges paid by users to prioritize their transactions and ensure they are processed quickly \citep{erdin2020bitcoin}. The distribution mechanism determines how the rewards issued by the system are allocated among miners after successfully solving the PoW puzzle. Typically, in Bitcoin, most of the mining work is done by so called \textit{pooled-mining} \citep{ren2019pooled}. In pooled mining, individuals collaborate by forming a mining pool, where they combine their computing power. This teamwork increases their chances of solving the PoW puzzle and receiving rewards more consistently. When a mining pool successfully solves a PoW puzzle, the distribution mechanism decides how the rewards are divided among the pool members. 

To better understand the interaction between Bitcoin's consensus and incentive mechanisms, consider an example of a transaction between two users. Suppose Alice wants to send 10 Bitcoins (BTC) to Bob. Alice's transaction data is broadcast across the Bitcoin network, entering a memory pool of unconfirmed transactions that await verification and inclusion in a new block. Miners continuously monitor this memory pool, selecting transactions to validate. They gather Alice’s transaction along with many others and aggregate these into a block. Each miner competes to solve the PoW puzzle associated with this block, which involves repeatedly attempting to find a solution by varying a small part of the block known as the nonce. When Jane, a miner, discovers a nonce that successfully solves the puzzle, she shares the solution with the network. The other miners then verify the solution, and once confirmed, the block is incorporated into the blockchain, provided that the majority of nodes approve it. As a result, the 10 BTC that Alice intended to send to Bob is successfully transferred, finalizing the transaction and securely recording it on the blockchain. Jane, the miner who solved the puzzle, receives a reward in the form of newly created Bitcoins, in addition to any transaction fees from the block.

Malicious parties can employ various strategies targeting consensus and incentive mechanisms to gain an unfair share of mining rewards, or manipulate transactions for personal financial gain. These attacks can take different forms, with some sticking to a single strategy, referred to as \textit{pure attacks} in this study. Alternatively, attackers might combine these pure attacks together or with other malicious and non-malicious strategies to enhance their effectiveness and profitability. We term these combined strategies as \textit{hybrid attacks}.

Under pure attacks, selfish mining-style attacks have been extensively explored in various studies \citep{eyal2018majority, nayak2016stubborn, sapirshtein2017optimal, negy2020selfish, grunspan2018profitability, yang2020assessing}. They enable a minority pool to earn more revenue than is equitable based on its total mining power \citep{eyal2018majority}. Bitcoin protocols prescribe that a miner who discovers blocks should immediately broadcast the valid blocks across the network. The miners who adhere to the Bitcoin protocols are called \textit{honest miners}. In the previous example, Jane is an honest miner as she published her block as soon as she discovered it. In contrast, a selfish mining pool keeps the newly mined blocks private and releases them strategically instead of broadcasting them immediately. Continuing from the previous example, let's assume Kevin is a rational miner leading a selfish mining pool that controls a large portion of the network's computational power. Jane, like other honest miners, works to confirm transactions and append new blocks to the blockchain, adhering to the rules of the network. Instead of immediately broadcasting newly mined blocks, Kevin’s pool withholds these blocks, creating a private chain, while Jane and the other honest miners work on the public chain. Suppose that, at one moment, the length of the honest chain is 1 and Kevin's private chain is 3, giving Kevin a lead of 2 blocks. If Jane successfully mines the next block, Kevin immediately publishes his private chain to the network. Since Bitcoin follows the rule of the longest chain, the network accepts Kevin’s chain, discarding the blocks that Jane and other honest miners had worked hard to add, thereby wasting Jane's computational efforts. Consequently, Kevin obtains the rewards of two blocks while Jane receives nothing. Selfish mining attacks present a significant threat to the fairness of the mining process, allowing attackers to earn rewards that exceed their fair share. Additionally, the resulting unfair distribution of rewards can lead some rational participants to engage in malicious behaviors \citep{eyal2018majority, wang2021forkdec, bastiaan2015preventing, sayeed2019assessing}. This, in turn, may result in a decrease in the number of honest miners, thereby weakening the network's security and creating opportunities for various types of attacks, particularly double-spending attacks \citep{sompolinsky2015secure}.

In a double-spending attack, the attacker spends the same cryptocurrency tokens more than once \citep{zhang2019double}. This allows the attacker to use the coins to purchase goods or services and then reverse the transaction while keeping both the goods/services and the coins. Essentially, this means obtaining the goods or services without spending any coins. Suppose the majority of the network's hash rate is under Kevin's pool's control. Because of this, the pool is able to mine blocks more quickly than Jane and other honest miners. Suppose, Kevin decides to use this advantage to double-spend his coins. For instance, Kevin buys a jet by spending a certain amount of his coins, and the transaction is broadcast to the network. Jane, the honest miner, includes Kevin's transaction in a block she successfully mines, extending the main chain and confirming the transaction. However, Kevin does not include his transaction in his private chain. Since Kevin’s mining pool can mine blocks faster, he is able to maintain a private chain that is longer than the public chain Jane is working on. While Jane and other honest miners contribute to extending the public chain, Kevin continues to mine additional blocks on his private chain, excluding the jet transaction. Once Kevin's private chain exceeds the length of the public chain, he publishes it to the network. As Kevin's chain is longer, the network accepts Kevin's chain, discarding the blocks mined by Jane, including the block with the jet transaction. As a result, the jet transaction is effectively reversed, allowing Kevin to reclaim the coins he spent on the jet while also receiving block rewards for his private chain. This successful double-spending attack not only allows Kevin to fraudulently regain his spent bitcoins but also reduces the trust among users and merchants and integrity of the entire blockchain network.

\subsection{Contributions} 
This paper provides several contributions to the field of blockchain security, with a focus on PoW-based blockchain networks:
\begin{enumerate}
    \item We provide a detailed examination of pure attacks on consensus and incentive mechanisms in PoW-based blockchain networks. Our analysis assesses the efficiency and profitability of these attacks when executed in isolation.
    
    \item This study investigates how pure attacks can be combined with other malicious and non-malicious strategies to form hybrid attacks. We analyze how these hybrid attack vectors create more sophisticated and effective attack strategies, enhancing attackers' success rates and profitability.
    
    \item Our analysis explores game theory-based approaches proposed by various authors to assess the dynamics and profitability of selfish mining and block withholding when two or more pools engage in these attacks against one another. By applying these models, we offer a quantitative understanding of the profitability of these pools in such adversarial environments.
    
    \item We propose a set of design guidelines to steer future research focused on preventing or mitigating the threats posed by the identified attack strategies.
\end{enumerate}

\subsection{Road map}
The remainder of the paper is structured as follows: Section~\ref{sec:ii} provides the blockchain preliminaries. Section~\ref{sec:iii} outlines the planning and execution of the conducted SLR. Section~\ref{sec:iv} presents the findings from the SLR, organized into pure attacks, hybrid attacks, and multiple pool attacks. Section~\ref{sec:v} provides the design guidelines to guide future research. Finally, Section~\ref{sec:vi} concludes the SLR.

\section{Preliminaries}
\label{sec:ii}

This section offers a brief overview of blockchain fundamentals, covering the PoW consensus algorithm, mining process, blockchain forks, and mining pools

\subsection{Proof of Work and Mining}

\begin{figure}[!t]
	\centering 
	\includegraphics[width=1.0\textwidth, keepaspectratio]{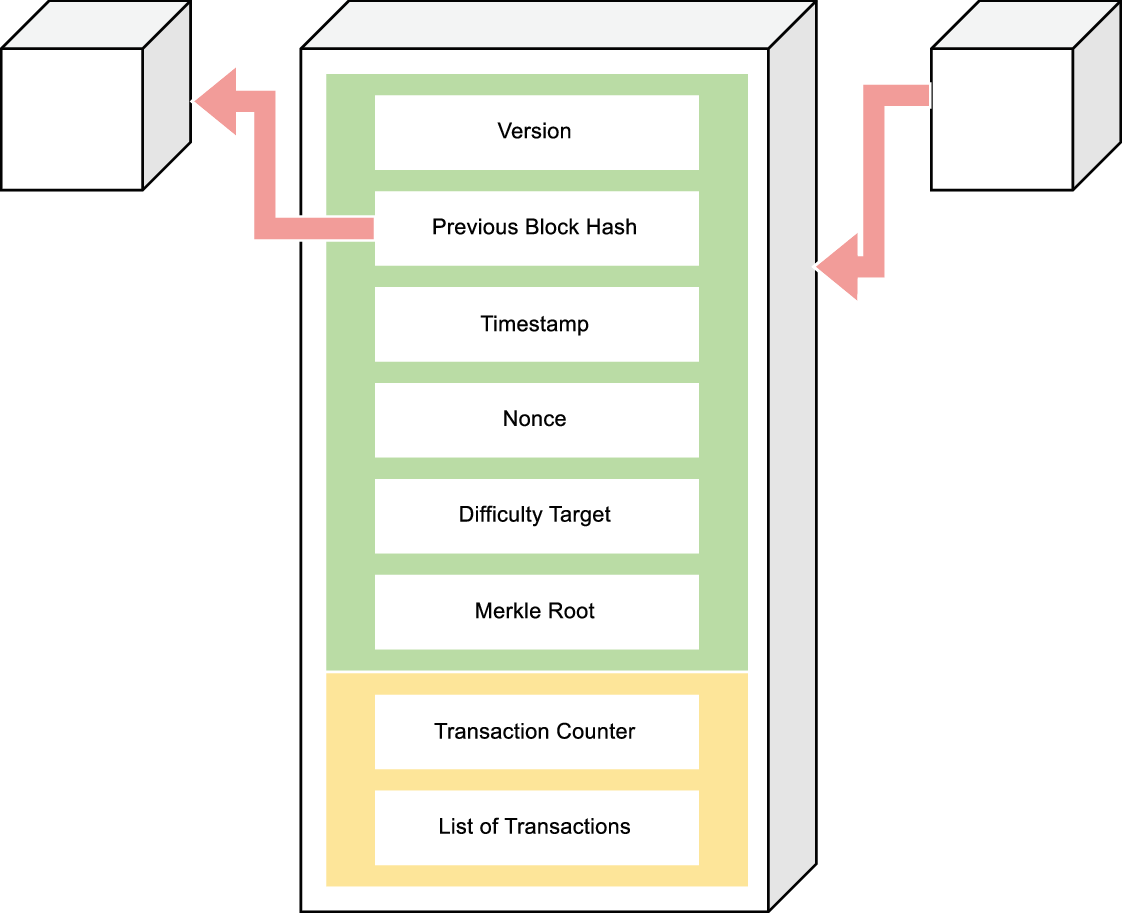}	
	\caption{Blockchain architecture} 
	\label{fig:blockchain_arch}%
\end{figure}

Blockchain is a distributed ledger technology that operates in a decentralized manner, enabling the secure and transparent recording of transactions across a network of computers \citep{sarmah2018understanding,aggarwal2021basics,sunyaev2020distributed}. At its essence, a blockchain consists of a series of blocks, each containing a collection of transactions. A block in the Bitcoin blockchain typically consists of components such as \textit{Block Header, Transactions, Block Size, Block Height, Block Hash}, and \textit{Block Reward} \citep{senmarti2015analysis}. The Block header contains metadata about the block. It includes \textit{Version, Previous Block Hash, Merkle Root} \footnote{The Merkle root of a block is a single cryptographic hash that uniquely represents the collective hash of all transactions in that block \citep{merkle1987digital}.}, \textit{Timestamp, Nonce,} and \textit{Difficulty Target}. The blocks are connected through cryptographic hashing, with each block referencing the hash of the preceding block in its header (Figure \ref{fig:blockchain_arch}), forming a chronological sequence that is highly resistant to tampering \citep{nakamoto2008bitcoin}.

The PoW algorithm functions as Bitcoin's consensus mechanism, facilitating agreement on the state of the blockchain and transaction validation \citep{nakamoto2008bitcoin,tschorsch2016bitcoin}.
It involves participants, referred to as miners, solving computationally demanding puzzles to generate new blocks and append them to the blockchain \citep{oyinloye2021blockchain}. To solve the PoW puzzle, a miner first selects a random nonce (a unique 32-bit number used only once) and constructs a Block header. Once the Block header has been constructed, the miner concatenates the Block header fields and hashes the concatenated string using a hashing algorithm (SHA-256 for Bitcoin \citep{nakamoto2008bitcoin}). If the generated hash is less than or equal to the current difficulty target, the miner has successfully solved the puzzle. This can be mathematically denoted as given below.

\begin{equation} \label{eq:hash_test}
    H\left(n,p,m\right)<D
\end{equation}
where \( n \) is the nonce value, \( p \) is the hash value of the previous block, \( m \) is the Merkle root of all the included transactions in the block, and \( D \) is the target.

The successful miner shares the new block with the network, along with the nonce and hash value. Other nodes in the network verify the block by independently hashing the block header with the nonce and comparing the resulting hash against the target. If the block is valid, it is accepted by the network and added to the blockchain. The miner who successfully mines a new block receives a specified number of newly created bitcoins, known as the block reward. Additionally, they may collect transaction fees related to the transactions included in the block.

The difficulty target is a 64-digit hexadecimal code that indicates how challenging it is to obtain a valid hash. An example difficulty target is given below. 

\begin{center}
\noindent \texttt{0000000000000000000abcd1234567890\allowbreak
fedcba09876543210abcdef123456789}
\end{center}

\begin{figure}[!t]
	\centering 
	\includegraphics[width=1.0\textwidth]{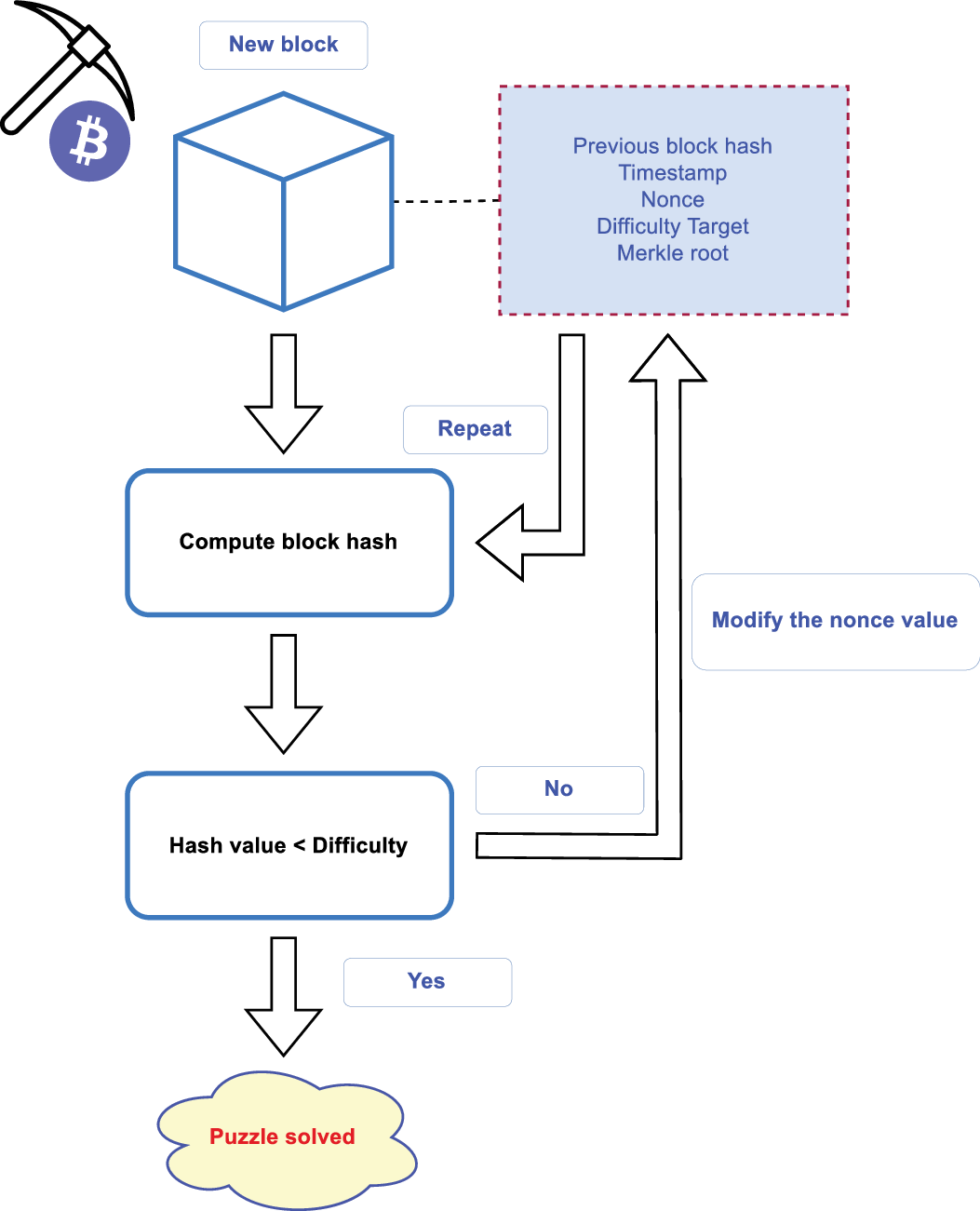}	
	\caption{PoW Algorithm}
	\label{fig:pow}
\end{figure}

The difficulty level is determined by the number of leading zeros in the difficulty target; the greater the number of leading zeros, the harder it is to find a valid hash. The difficulty target is adjusted periodically (e.g., every 2016 blocks in Bitcoin \citep{nakamoto2008bitcoin}) to ensure blocks are mined at a consistent rate (approximately every 10 minutes for Bitcoin \citep{kraft2016difficulty}). The target hash given above has a large number of leading zeros, indicating a high level of difficulty. This process is known as “proof of work" because the miner has demonstrated that they have expended computational effort (work) to find a valid hash. The PoW algorithm can be summarized as shown in Figure \ref{fig:pow}.

\subsection{Blockchain Forks}
A blockchain fork occurs when the main chain splits into two separate branches \citep{neudecker2019short}. This situation typically arises when two miners simultaneously discover and broadcast different valid blocks that reference the same preceding block. Each branch of the blockchain now contains a different valid block at the same height. When both blocks are broadcast to the network, nodes receive and propagate both versions. Consequently, different parts of the network may temporarily see different versions of the blockchain, leading to a brief period of uncertainty about which block is the "correct" one. During this time, miners and nodes will continue to build upon the block they received first, effectively extending their respective branches of the blockchain. This creates a competitive environment where miners compete to discover the next block. The branch that expands the fastest—meaning the one with the most blocks added—ultimately becomes the dominant chain. According to the blockchain's consensus rules, the network will recognize the longer chain as the valid one, while the shorter branch, which contains fewer blocks, is discarded. Blocks that were part of the shorter branch are considered orphaned or stale. Transactions from these orphaned blocks are returned to the memory pool and can be included in subsequent blocks. This process ensures that the blockchain eventually converges back to a single, unified chain, maintaining the network's integrity and consensus.

\subsection{Mining Pools}
\label{sec:ii-c}

The current hash rate of modern mining operations exceeds \( 7 \times 10^{19} \) hashes per second \citep{gao2019power}. As a result, the likelihood of an individual miner successfully discovering a block is extremely low. To address this, miners join mining pools to achieve a steadier income and reduced variance in rewards \citep{eyal2018majority}. Mining pools provide miners with a higher probability of mining blocks and earning rewards, thereby reducing the financial risks associated with solo mining by increasing their chances of earning rewards more consistently \citep{romiti2019deep}. Nowadays, more than 90\% of cryptocurrency mining is conducted through pooled mining \citep{haghighat2019block}.

In a mining pool, miners collaborate by pooling their resources to jointly work on block discovery and share the resulting rewards. In a typical setup, a pool operator oversees the management of mining, coordinating the pool's activities. The operator sets up and maintains the pool’s server, monitors its performance, distributes the rewards among the participants, and may charge a fee for their services. Individual miners join the pool by connecting their mining hardware to the pool's server, allowing their computational power to be combined with that of other miners. The pool operator assigns smaller, manageable tasks to each miner, and miners contribute to the pool's computational effort by finding and submitting shares. When a miner finds a block that produces a hash starting with a considerable number of zeros, they submit this hash to the pool manager as a share. Each hash attempt has a probability of \( \frac{1}{2^{32}} \) of resulting in a share. Solving shares involves the same process as mining a block, but shares are solutions that meet a lower difficulty target set by the pool, rather than the full network difficulty. The pool manager verifies the share submitted by the miner to ensure it meets the pool's difficulty target. When a miner in the pool discovers a share that satisfies the network's difficulty target, it is submitted as a valid block to the blockchain network. Upon finding a valid solution that meets the network's difficulty, Bitcoin's incentive mechanism decides how the rewards are distributed among the miners involved in the pool. These distribution mechanisms, also known as \textit{payout} or \textit{reward schemes}, vary in structure and impact. Below are some of the most common payout schemes:

\begin{itemize}

\item Pay-per-Share (PPS): A simple and fixed payout scheme compensates miners for each valid share they contribute to the pool, regardless of whether a block is discovered. Although this method ensures that miners receive payment, it presents a financial risk to the pool, as it must pay miners even when no block is successfully mined.

\item Proportional (PROP): Unlike PPS, this scheme rewards miners only after a block is found. The payout corresponds to the miner's share of the pool's computational power. For example, if a miner contributes 5\% of the pool's total power, they would receive 5\% of the block reward.

\item Pay-per-Last-N-Shares (PPLNS): Like PROP, PPLNS rewards miners based on the number of shares they have submitted in the most recent 'N' blocks instead of their total contribution. This approach helps prevent pool hopping and encourages loyalty, as miners earn smaller rewards initially, which gradually increase over time.

\item Dynamic Pay-per-Last-N-Shares (DPPLNS): This scheme operates like PPLNS but adds a dynamic element, adjusting the number of blocks considered for share calculation, which can provide a more responsive reward system.

\item PPS+: A hybrid of PPS and PPLNS, where the base block reward is distributed under the PPS method, and transaction fees are paid out using the PPLNS scheme. Miners may not always achieve the expected results when switching pools during high transaction fee periods.

\item Full Pay-per-Share (FPPS): In this scheme, both the base block reward and transaction fees are paid out under the PPS method, offering miners consistent and predictable payouts.

\item SOLO: In this scheme, a miner is only rewarded if they personally find a block, offering potentially high but infrequent payouts.

\item PPLNS+: A variant of PPLNS that averages rewards based on block output. Like PPLNS, initial rewards are modest, but they increase steadily over time, typically reaching the expected value within 24 hours.

\item Double Geometric Method (DGM): A hybrid between the Geometric and PPLNS methods, where share value is calculated at the start of each round, with the pool buffering rewards during short rounds to pay out during longer ones.

\item Pay-per-Last-N-Timeframes (PPLNT): Similar to PPLNS, this method averages rewards over specific timeframes (e.g., 30 minutes) instead of block search rounds.

\item RBPPS: This scheme resembles PROP, but the rewards depend on the duration it takes the pool to discover a block since the previous block was mined, rather than the time the network takes to mine that block.

\item Pay-per-Loyal-Time-Share (PPLTS): A combination of PROP and PPLNS, this scheme rewards consistent hash rates and "punishes" sudden spikes in computational power. Miners who regularly contribute to the pool see better rewards.

\item Predictable SOLO (PSOLO2): Here, the miner is rewarded once their total contribution matches the current network difficulty, providing a predictable payout system for solo miners.

\end{itemize}

\section{Methodology}
\label{sec:iii}

The study utilized a Systematic Literature Review (SLR) methodology to address the research questions detailed in Section~\ref{sec:iii-a} In accordance with the updated 2020 Preferred Reporting Items for Systematic Reviews (PRISMA) guidelines \citep{page2021prisma}. The selection of studies for our SLR focusing on attacks targeting consensus and incentive mechanisms in PoW-based blockchain networks involves a rigorous, transparent, and methodical process to ensure the inclusion of relevant, high-quality research. The selection process consisted of the following steps:

\subsection{Formulation of research questions}
\label{sec:iii-a}
To guide our investigation into attacks on consensus and incentive mechanisms, we formulated the following research questions (RQs):
\begin{center}
\begin{itemize}
    \item RQ1: \textit{How do pure attacks perform in isolation within PoW-based blockchain systems and what are their effectiveness and profitability?}
    \item RQ2: \textit{What are the impacts on their effectiveness and profitability when these attacks are combined, either with each other or with other malicious or non-malicious strategies?}
    \item RQ3: \textit{How does the competition between multiple mining pools deploying attacks like selfish mining and block withholding against each other affect the profitability of each pool?}     
\end{itemize}
\end{center}

By exploring these research questions, this paper seeks to offer an in-depth analysis of the different attacks directed at consensus and incentive mechanisms within PoW-based blockchain networks. This investigation will enhance the understanding of these attacks  and guide future research directions in improving resilience of blockchain networks.

\subsection{Data Sources and Search Strategies}
The SLR conducted in this study involved an in-depth exploration of published articles spanning the period from 2018 to 2024 across a wider range of electronic databases: Scopus, IEEE Xplore Digital Library, Springer Link and Google Scholar. These databases were selected due to their globally acknowledged impact indices, which encompass a wide array of peer-reviewed scientific and scholarly literature from various scientific domains and disciplines worldwide.
We determined search strings by identifying associated key terms within consensus and incentive attacks. This was based on our subject knowledge and previous most-cited research papers and journals. 
We established two generic search terms in association with informatics and employed Boolean operators as follows to encompass all literature focusing on bitcoin-like PoW-based blockchain systems.
\begin{center}
\textit{(blockchain AND bitcoin)}
\end{center}

Drawing upon subject knowledge, we identified various types of attacks targeting consensus and incentive mechanisms in PoW-based blockchain networks. Subsequently, we formulated search terms tailored to capture these attacks effectively as given below.
\begin{center}
(\textit{selfish mining}, \textit{51\% attack
}, \textit{pool hopping
}, \textit{stubborn mining 
}, \textit{block discarding
}, \textit{block withholding
}, \textit{honest mining
}, \textit{mining attack
}, \textit{faw attack}, \textit{double spending})
\end{center}

Subsequently, we combined these key values into the generic query and executed iterative queries across digital libraries for each attack. A few example search queries are given below.
\begin{center}
\begin{itemize}
    \item Query1: \textit{("blockchain" AND "bitcoin") AND ("mining attack")}
    \item Query2: \textit{(“blockchain” AND “bitcoin”) AND (“selfish mining”)}
    \item Query3: \textit{(“blockchain” AND “bitcoin”) AND (“51\% attack”)}    
\end{itemize}
\end{center}

\subsection{Selection of studies}
These inclusion criteria help ensure that our SLR focuses on relevant, high-quality research that contributes meaningfully to understanding attacks on consensus and incentive mechanisms of PoW-based blockchain networks.
\begin{itemize}
\item Empirical studies employing quantitative, qualitative, or mixed-methods research designs, including experimental studies, case studies, surveys, and observational studies.

\item Articles that specifically focus on the development, implementation, assessment, or implications of various attacks on the consensus and incentive mechanisms of PoW-based blockchain networks.
\item Only the articles published in English are considered.
\item Articles with a clear description of research methods, data collection techniques, analysis procedures, and consideration of potential sources of bias.
\item Articles available through academic databases, institutional repositories, or other accessible sources for data extraction and analysis.
\end{itemize}

We established these inclusion criteria to ensure our SLR includes only the most relevant and high-quality studies. This strategy enhances the reliability and validity of our findings, providing a solid basis for understanding the various attacks on consensus and incentive mechanisms in PoW-based blockchain networks.

\subsection{Screening Process}

The screening process for our SLR utilizes a systematic and thorough method to guarantee the inclusion of relevant, high-quality studies. It comprises two primary stages: initial screening and full-text screening. This process helps in efficiently filtering out irrelevant studies and selecting those that meet the predefined inclusion criteria.

\subsubsection{Initial Screening}

The initial screening phase entails an initial review of the titles and abstracts of all studies obtained from our search strategy. The primary goal is to quickly assess each study's relevance to the research questions and eliminate those that do not fulfil the fundamental inclusion criteria. During this phase, the titles and abstracts of 518 publications were reviewed by two independent reviewers to assess their relevance. The reviewers then applied predefined inclusion criteria, focusing on factors such as relevance to attacks on consensus and incentive mechanisms of PoW-based blockchain networks, publication type, clarity, and language. Every decision to include or exclude a study was carefully recorded, with justifications provided for any exclusions when relevant. When disagreements occurred between the two reviewers, they were resolved through discussion or by involving a third reviewer. This transparent process ensured that only studies meeting the necessary criteria progressed to the next stage of screening. At the end of this process a sum of 71 publications were submitted to the next phase.

\subsubsection{Full-Text Screening}

\begin{figure}[!t]
    \centering
    \includegraphics[width=1.0\textwidth, keepaspectratio]{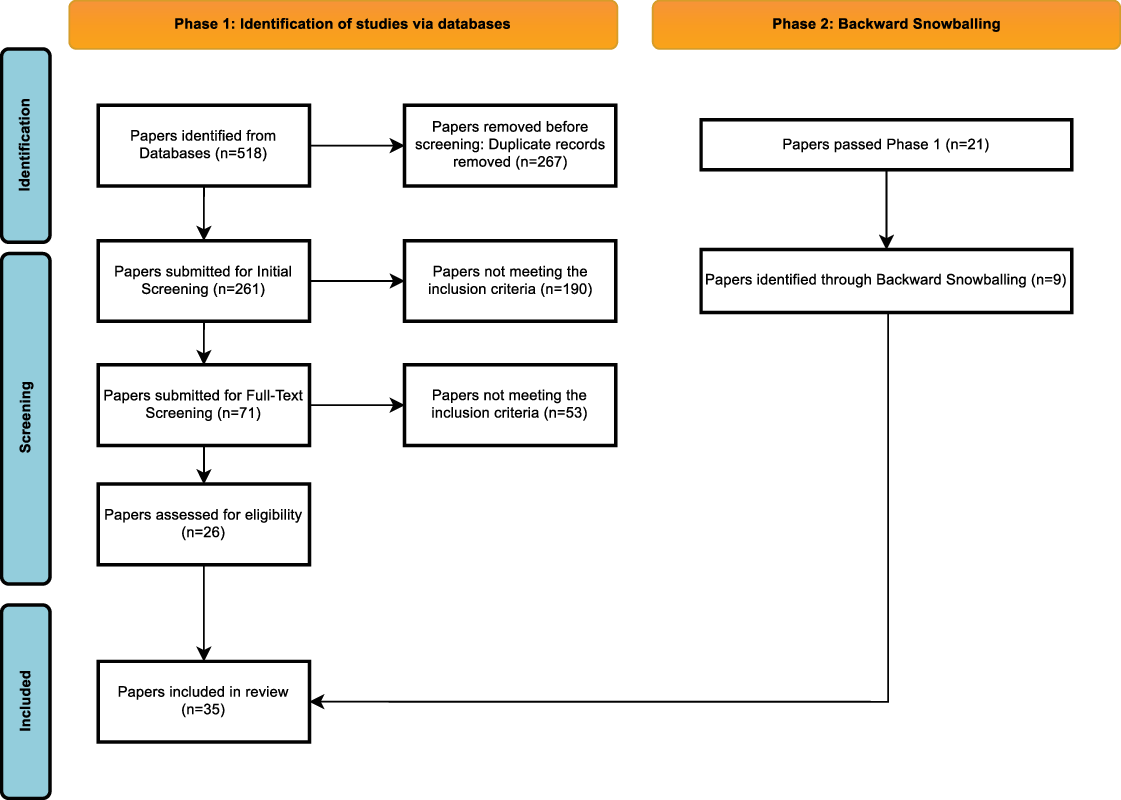}
    \caption{PRISMA flow diagram}
    \label{fig:prisma_flow}
\end{figure}

During the full-text screening phase, the full texts of the studies that passed the initial screening were thoroughly reviewed to ensure they met all inclusion criteria. Each study was evaluated in detail, focusing on the methodology, relevance to attacks on consensus and incentive mechanisms of PoW-based blockchain networks, and the quality of evidence provided. 

Decisions regarding the inclusion or exclusion of studies were thoroughly documented, with clear explanations provided for any exclusions. Disagreements among reviewers were addressed through discussions or by seeking the input of a third reviewer. This detailed review process ensured that only the most relevant and high-quality studies made it into the final analysis. At the end of this process, a total of 35 publications were selected for the SLR (Table \ref{tab:publications}). A PRISMA flow diagram was employed to illustrate the study selection process, detailing the number of studies that were identified, screened, deemed eligible, and ultimately included in the review (Figure \ref{fig:prisma_flow}).

\begin{table}[!t]
    \centering
    \resizebox{\textwidth}{!}{
        \begin{tabular}{ll}
            \toprule
            \textbf{Publication} & \textbf{Attack} \\
            \midrule
            \citep{eyal2018majority,grunspan2018profitability,courtois2014subversive,wright2017fallacy} & Selfish mining \\
            \citep{sapirshtein2017optimal} & Optimal selfish mining \\
            \citep{wang2021blockchain} & Optimal selfish mining with Reinforcement Learning \\
            \citep{gervais2016security} & Optimal selfish mining and double-spending with eclipse attacks \\
            \citep{negy2020selfish} & Intermittent selfish mining \\
            \citep{nayak2016stubborn} & Stubborn mining \\
            \citep{iqbal2021exploring} & Sybil and double-spending, Eclipse based double-spending, BGP Hijacking for double-spending \\
            \citep{heilman2015eclipse} & Selfish mining and double-spending with eclipse attacks \\
            \citep{finney_attack} & Finney attack \\
            \citep{race_attack} & Race attack \\
            \citep{vector76_attack, vector76_attack_2} & Vector-76 attack \\
            \citep{br_attack} & Blockchain reorganization attack \\   
            \citep{liu2018strategy,zhang2020analysing,bai2023blockchain} & Selfish mining (Multiple attackers) \\
            \citep{bai2019deep} & Selfish mining (3-player game)  \\
            \citep{li2021semi} & Semi selfish mining \\
            \citep{sompolinsky2015secure} & Double spending \\
            \citep{bastiaan2015preventing,sayeed2019assessing} & 51\% attack \\
            \citep{belotti2018bitcoin} & Pool Hopping attack \\
            \citep{bahack2013theoretical} & Block Discarding attack, Difficulty Raising attack \\
            \citep{rosenfeld2011analysis} & Block Withholding attack (single pool) \\
            \citep{eskandari2020sok} & Front-running attack \\
            \citep{li2020mining,wu2019equilibrium,qin2020optimal} & Block withholding attack (dual mining pool) \\
            \citep{haghighat2019block} & Block withholding attack (multiple pools) \\
            \citep{kwon2017selfish} & Fork After Withholding attack (FAW) \\
            \citep{wang2023efaw} & FAW attack with eclipse attack \\
            \citep{gao2019power} & Power Adjusting Withholding, Bribery Selfish Mining attacks \\
            \bottomrule
        \end{tabular}
    }
    \caption{List of 35 publications included in the review}
    \label{tab:publications}
\end{table}

\section{Results}
\label{sec:iv}

In this section, we present the findings of our SLR, organized to address our three primary research questions. Section~\ref{sec:iv-a} focuses on pure attacks, directly mapping to RQ1, where we evaluate the effectiveness and profitability of individual attack strategies within PoW-based blockchain systems. Section~\ref{sec:iv-b} delves into hybrid attacks, aligning with RQ2, as we investigate how the combination of various attack methods and strategies influences their overall effectiveness and profitability. Finally, Section~\ref{sec:iv-c} examines multiple pool attack analysis, corresponding to RQ3, where we analyze the competitive dynamics between mining pools that deploy attacks like selfish mining and block withholding against one another, highlighting the impacts on profitability in this adversarial environment. 

\subsection{Pure attacks analysis}
\label{sec:iv-a}

In our study, we define pure attacks as individual, discrete strategies employed by attackers to exploit vulnerabilities in consensus and incentive mechanisms of PoW-based blockchain systems. These attacks are executed independently, without the need to combine them with other strategies. In this section, we analyze the mechanism behind these strategies and evaluate their effectiveness and profitability.

\subsubsection{Selfish mining} 

\citet{eyal2018majority} proposed a strategy known as selfish mining, which allows a minority pool to earn more revenue than its share of the overall mining power. The concept behind selfish mining involves the pool intentionally forking the main chain by concealing its mined blocks from public view. As a result, the pool mines on its own private branch while the honest network continues to mine on the public chain. By consistently adding more blocks, the pool creates a significant lead over the public chain. When the public chain gets close to matching the length of the pool's private branch, the pool releases blocks from its private chain to the public, effectively discarding the blocks mined by the honest network. This compels honest miners, who are following the Bitcoin protocol, to waste resources on solving cryptographic puzzles that yield no real benefit. \citet{eyal2018majority} outlined the strategy of selfish mining by assuming that miners are categorized into two groups: an adversarial minority pool that employs the selfish mining approach and possesses a fraction \(\alpha\) of the network's total computational power, and honest miners, who form the majority and follow the standard Bitcoin protocol, controlling the remaining portion \(1-\alpha\) of the network's computational power. They introduced a parameter known as communication capability denoted as \(\gamma\) , which indicates the fraction of honest miners who selects pool's branch during a blockchain fork, while the remaining fraction \(1 - \gamma\) mines on the other branch. This occurs because when both the pool and an honest miner publish their newly mined blocks around the same time, not all miners in the network receive the notification immediately due to communication and propagation delays. As a result, only a portion of honest miners become aware of the new block. Their findings suggest that a selfish mining attack becomes financially viable when the pool possesses a minimum of 33\% of the total mining power in the network. This is known as the \textit{profitability threshold}.

Studies \cite{sapirshtein2017optimal,gervais2016security} show that selfish mining is not profitable when nodes do not employ a large enough share of the computing resources or due to communication limitations of the pool. Selfish mining is more profitable than honest mining if the mining pool can propagate its newly mined block quickly to all honest miners. In particular, when \(\gamma=1\) the profitable threshold significantly decreases to zero because of the pool's ability to rapidly propagate its block ensures that honest miners will mine on the pool's block. Conversely, when \(\gamma=0\), the profitability threshold rises to \( \frac{1}{3} \), as honest miners consistently publish their newly mined blocks before the selfish miners can propagate theirs. 

Another important consideration is that the selfish mining strategy provides no benefit to the adversary until a difficulty adjustment \citep{grunspan2018profitability}. When the adversary withholds blocks and releases them selectively, they create irregularities in block discovery times, thereby reducing the block discovery rate. Upon reaching the difficulty adjustment period, the network lowers the difficulty in response to the perceived slower discovery rate. As a result, the adversary can now mine at a reduced difficulty level, enabling them to gain a higher proportion of the total block rewards with less computational effort. If the network difficulty were to remain constant, the immediate rewards from this strategy might not be sufficient to offset the associated mining costs, such as electricity and hardware expenses.

Some authors have denied the profitability of selfish mining based on incorrect models of the Bitcoin protocol and how pools work \citep{grunspan2018profitability,courtois2014subversive,wright2017fallacy}. They contend that selfish mining is unprofitable, due to the time dedicated to forking blocks ultimately slows down the growth rate of the main chain. For instance, if an attacker successfully generates \(x\) blocks during an attack and the honest network produces \(y\) blocks, then all of the blocks mined by the attacker will ultimately be part of the main chain. In comparison, the honest network will have only \(y-x\) of its blocks included. Thus, the blockchain ultimately grows only by \(y=(x+y-x)\) blocks. The attacker strategically publishes each block they mine whenever an honest miner discovers a new block, effectively substituting their own block for the competing one in the chain. As a result, for every \textit{x} blocks mined by the attacker, an equivalent number of blocks from the honest network are discarded and substituted with the blocks mined by the attacker. Consequently, this reduces the average rate with which the blockchain grows by a factor \(1-p\) compared to the normal rate, where \(p\) is the fraction of the network's mining power controlled by the attacker. Based on this argument, because the blockchain grows more slowly due to the attacker's blocks replacing honest blocks, the relative revenue from selfish mining is reduced. As this reduces the revenue per unit of time, the strategy will be less effective. Furthermore, critics have pointed out that for selfish mining to be profitable, long duration attacks are required where the attack extends beyond a single adjustment period. The critics also believe that this requirement of long-duration attacks reduces the feasibility and profitability of the selfish mining strategy  All these arguments proved to be false in a later study \citep{negy2020selfish}.

\subsubsection{Stubborn Mining} 

\citet{nayak2016stubborn} introduced a set of mining strategies known as stubborn mining, which not only extend the selfish mining but also enable miners to achieve greater profits. The core idea behind stubborn mining strategies is that attackers can frequently achieve higher profits by continuing to mine on their private chain, even when their private chain lags behind the honest chain. This contrasts with the selfish mining strategy, where the selfish miner withholds their mined blocks only when they are ahead and adopts to the honest chain when they fall behind. In their study, three stubborn mining strategies were presented: Lead Stubborn (L-stubborn), Equal Stubborn (F-stubborn), and Trail Stubborn (\( T_j \)-stubborn) mining. Each strategy involves slightly different behaviors from the attacker.

In the L-stubborn strategy, an attacker risks their mined blocks by selectively releasing them, even when maintaining a significant lead over the public chain. When the honest network finds a block, the attacker reveals only enough blocks to match the public chain’s length, rather than disclosing their entire private chain, as seen in selfish mining. This results in a fork in the blockchain, with a fraction \(\gamma\) of honest miners working on the attacker’s chain and the remainder \(1 - \gamma\) continuing on the public chain. This strategy carries the risk of losing the private chain if the public chain ultimately becomes the accepted chain.

In the F-Stubborn strategy, the attacker refrains from revealing their next block when the blockchain encounters an equal-length fork. Rather than disclosing their block to match the public chain, the attacker retains it privately and continues mining on their own chain. This approach contrasts with selfish mining, in which an attacker would usually release their next block to match the length of the public chain.

The idea behind \( T_j \)-stubborn strategy is that the attacker persists in mining on their private chain, even when it lags behind the public chain, aiming to eventually close the gap and overtake it. Under \( T_j \)-stubborn strategy, \citet{nayak2016stubborn} presented a family of "trail stubborn" strategies characterized by the parameter \(j\). This parameter determines that the attacker will abandon their private chain if it falls behind by \(j+1\) blocks compared to the public chain.

\citet{nayak2016stubborn} evaluated the profitability of honest mining, selfish mining, and their stubborn mining strategies across various combinations of \(\alpha\) and \(\gamma\) values. Their results indicate that stubborn mining techniques can yield profits up to 25\% higher than selfish mining, even without hybrid attacks involving network-level exploits, across a range of plausible \(\alpha\) and \(\gamma\) values.

\subsubsection{Optimal selfish mining}

\citet{sapirshtein2017optimal}, used a general Markov Decision Process (MDP) \citep{puterman1990markov} model, to represent the mining process. Solving this MDP yields an optimal selfish mining (OSM) strategy, enabling even smaller mining pools to achieve higher rewards than through standard selfish mining. They initially formulated the mining process as a single-player decision problem with a non-linear objective function (Equation \ref{eq:expected_gain}). 

\begin{equation} \label{eq:expected_gain}
    REV = \frac{E\left[\sum_{t=1}^{T} r_t^1 (\pi)\right]}{E\left[\sum_{t=1}^{T} r_t^1 (\pi) + \sum_{t=1}^{T} r_t^2 (\pi)\right]}
\end{equation}

where \( r_t^1 (\pi) \), \( r_t^2 (\pi) \) is the immediate reward issued in the block interval \( t \) under the action defined by policy , and \( T \) is the size of the observing window. The action space includes four actions: \textit{Adopt}, \textit{Override}, \textit{Match}, and \textit{Wait}. Each state is represented by a three-element tuple \( (l^{(a)}, l^{(h)}, \textit{fork}) \), where \(l^{(a)}\) and \(l^{(h)}\) denote the lengths of the adversary's and the honest chain, respectively, since the last fork. The \textit{fork} element indicates one of three states: \textit{relevant}, \textit{irrelevant}, or \textit{active}. To obtain an optimal mining policy, the study addressed the the non-linear objective function by initially converting the problem into a series of MDPs with linear objectives. Subsequently, they employed a standard MDP solver alongside a numerical search across these MDPs to derive the optimal selfish mining policy. Their simulations reveals that an attacker can achieve a greater portion of the rewards by employing a lower profit threshold of 23.21\%. 

However, this approach faced limitations due to its model-based nature. It requires knowledge of network parameters such as the attacker’s computational power (\(\alpha\)) and communication capability (\(\gamma\)). In real blockchain networks, these values are difficult to determine due to their variability \citep{wang2021blockchain}. Moreover, the model does not account for various blockchain characteristics such as stale block rates, confirmation times, and real-world parameters like network delays \citep{gervais2016security}. Consequently, there is a substantial gap between this model and real-world blockchain networks.

\subsubsection{Optimal selfish mining with Reinforcement Learning}

In contrast to the model-based approach presented by \citet{sapirshtein2017optimal}, \citet{wang2021blockchain} proposed a model-free, Reinforcement Learning (RL) \citep{li2017deep,lapan2020deep} based approach which allows an RL agent to dynamically learn a mining strategy with performance approaching that of the optimal mining strategy. For their work, they adopted the MDP mining model proposed by \citet{sapirshtein2017optimal} with tabular Q-Learning \citep{watkins1992q} algorithm to derive the optimal mining strategy. In their research, they employed a refined version of the Q-Learning algorithm, termed the \textit{Multidimensional Q-Learning algorithm}, owing to the non-linearity inherent in the objective function (Equation \ref{eq:expected_gain}). Leveraging this multidimensional Q-Learning algorithm, they successfully optimized the non-linear objective function to achieve the optimal mining strategy. 

However, this approach faced limitations in its applicability to a real blockchain environment due to two primary reasons. Firstly, it relied on the MDP model proposed by \citet{sapirshtein2017optimal}, which lacked consideration for real-world blockchain parameters like stale block rates, eclipsed attacks, propagation parameters, among others. Secondly, in their study, they employed a tabular Q-Learning algorithm, which is highly inefficient for handling realistic blockchain environments with large state spaces. Indeed, this directly impacts the convergence of the algorithm. If the algorithm requires a substantial amount of time to discover the optimal policy, it becomes economically unviable for miners. This is because prolonged computation time translates to increased expenses for hardware and computing power that the miners has to expend, reducing the economic feasibility of mining operations.

\subsubsection{Double-spending}

\citet{sompolinsky2015secure} show that malicious entities possess the capability to disrupt the synchronization of the ledger across multiple nodes, facilitating a form of attack known as the double-spending attack. In a double-spending attack, the attacker spends the same cryptocurrency tokens more than once. There are many variants of double-spending attacks.

The Race attack exploits the traders and merchants who accept payments immediately upon receiving "0-unconfirmed" transactions \citep{aggarwal2021attacks,race_attack}. A 0-confirmation transaction is simply a  transaction that has been broadcast to the network but has not yet been included in a block and added to the blockchain. These transactions are risky because they are susceptible to being reversed or replaced as they have not yet been confirmed. The Race attack involves the following steps.

\begin{enumerate}
    \item The attacker, controlling addresses \( X \) and \( Y \), initiates a transaction \(T_1\), from address \( X \) to a merchant address \( Z \). \(T_1\) is broadcast to the network and accepted by the merchant as a payment.
    While doing so, the attacker creates a conflicting transaction \(T_2\) from address \(X\) to \(Y\) spending the same coins. The attacker’s goal is to have \(T_2\) confirmed in a block before \(T_1\).
   
    \item Different nodes in the network might receive these transactions at different times because of the propagation delays in the network. If \(T_1\) reaches the merchant first, they provide the goods before the network has fully propagated \(T_2\). 

    \item If T2 is included in a block and validated by the network before T1, it effectively cancels out Transaction T1.
    
\end{enumerate}

Hence, the merchant, who has already provided the goods with the expectation of receiving the payment from \(T_1\), ends up without the payment. On the other hand, the attacker has effectively obtained the goods or services for free from the merchant.

The Finney attack, named after Hal Finney, describes a variation of the double-spend attack targeting merchants who accept payments with 0-confirmation transactions \citep{finney_attack}. This attack involves the following steps.

\begin{enumerate}
    \item The attacker, possessing some mining capability, occasionally mines blocks. In each block created, they include a transaction that transfers funds from address \( X \) to address \( Y \), both of which they control. The attacker keeps the mined block private without immediately broadcasting it.
    \item The attacker makes a purchase from a merchant and makes a payment to merchant's address \( Z \) from address \( X \). 
    \item The merchant proceeds with fulfilling the order and transferring the goods.
    \item After the merchant have completed the transaction, the attacker then broadcasts the block containing the transaction from \( X \) to \( Y \). Since this block was mined before the transaction to the merchant's address \( Z \), the blockchain will accept this version, causing the transaction to the address \( Z \) to be  rolled back.
\end{enumerate}

Unlike Finney attack and Race attack, the Blockchain Reorganization (BR) attack, also known as the Alternative History attack, is a different type of double-spending that can reverse transactions even after several confirmations \citep{br_attack}. This attack demands the attacker to have a significant amount of mining power and resources to be able to mine faster than the network. It involves the following steps:
\begin{enumerate}
    \item The attacker, controlling addresses \( X \) and \( Y \), initiates a transaction \(T_1\) from address \( X \) to a merchant address \( Z \). This transaction is broadcast to the network and accepted by the merchant as a payment.

    \item While doing so, the attacker privately mines an alternative blockchain fork. This fork includes a fraudulent transaction \(T_2\) where the attacker sends the same coins from address \( X \) to address \( Y \), thereby double-spending the same coins.

    \item The merchant waits for \( n \) confirmations of the transaction \(T_1\) before delivering the product or service. 
    
    \item If the attacker manages to mine more than \( n \) blocks on their private fork at this point, they can broadcast this longer fork to the network. Since this fork is longer than the public blockchain, it becomes accepted. This results in the transaction \(T_1\) invalidated, and the attacker reclaims their coins.
\end{enumerate}

When launching a BR attack, the attacker takes a risk: if they fail to mine a longer fork than the public chain, the attack fails, wasting their computational effort. 

In a 51\% attack, an attacker who commands 51\% of the hashing power of the entire network engages in double-spending \citep{bastiaan2015preventing,sayeed2019assessing}. Having a majority of hash power, enables the attacker to mine blocks at a faster rate than other miners. A 51\% attack includes the following steps.

\begin{enumerate}
    \item The attacker, controlling addresses \( X \) and \( Y \), initiates a transaction \(T_1\) from address \( X \) to a merchant address \( Z \). This transaction is broadcast to the network and is included in the public blockchain.
    \item The attacker begins to mine blocks privately, without broadcasting them across the network, thereby maintaining their own private chain. In their private chain, the attacker creates a conflicting transaction \(T_2\) that spends the same coins as \(T_1\) but sends it to the address \(Y\) controlled by the attacker.
    \item As the attacker possesses a majority of hash power, they can mine blocks on the private chain faster than the network. This results in the private chain becoming longer than the public chain.
    \item Once the private chain is longer than the public chain, the attacker broadcasts this longer chain to the network. 
    \item The network accepts the attacker's private chain as it is longer, thereby discarding the public chain. This results in the transaction \(T_1\) invalidated, and the attacker reclaims their coins.
\end{enumerate}

\citet{nakamoto2008bitcoin} presented that if an attacker controls less than 50\% of the computational power within the network, the likelihood of successful double-spend attacks decreases exponentially over time. Typically, merchants and exchanges require multiple confirmations to consider a transaction final and irreversible. The analysis done by \citet{sompolinsky2015secure} indicates that the success of double-spending attack does not just depend on how many blocks have been added after the transaction (confirmations) but also on how long it takes for the transaction to be added to the blockchain. Given that blocks are typically generated approximately once every 10 minutes in Bitcoin, there exists a significant delay before a specific transaction is included into the blockchain.

\subsubsection{Pool Hopping attack}
The Pool Hopping attack \citep{belotti2018bitcoin} involves miners carefully timing when they join and leave a mining pool to increase their mining reward. In this strategy, miners direct their computational resources to the pool when it's most advantageous, then pull out and move their resources elsewhere when the potential rewards decrease. By doing so, these miners manage to earn rewards that are higher than what would be fair based on their actual contribution to the pool’s overall computational power, resulting in smaller rewards for other miners. Pool hopping is particularly effective in pools that use the proportional reward system, which is popular because of its simplicity. In this system, the value of a share submitted at any given moment is influenced by the total number of shares submitted since the last block was found. It means during a "short mining round", fewer shares are generated, so each share holds a relatively higher value in the proportional reward system. Miners participating in a short round may earn a higher reward per share because the total number of shares submitted is low, leading to a larger portion of the block reward being distributed to each participating miner. In contrast, a "long mining round" occurs when it takes more time to discover the next block, resulting in a larger number of shares being submitted. This makes the value of each share less worth. A rational miner can exploit this situation by strategically submitting shares to the pool during shorter rounds and redirect their efforts elsewhere during longer rounds. As a result, the rational miner will leave the pool when it becomes less rewarding and redirect their efforts to another pool with better rewards. This strategic shifting of resources allows them to increase their mining reward at the expense of miners who consistently contribute to the pool.
This is because, since pool hoppers get a disproportionate share of the rewards, while consistent miners who contribute throughout the round receive less than their fair share. This unfair distribution discourages regular miners from staying in the pool, as they end up earning less rewards than what they would in a fair system. Moreover, if more miners engage in pool hopping, the pool becomes less predictable in terms of reward distribution. This inconsistency can lead to a loss of trust among the pool's regular miners, causing them to leave the pool in search of more stable and fair alternatives. Over time, this can reduce the pool's overall hash rate, making it less competitive and less likely to find blocks, further reducing rewards for regular miners who stay.

\subsubsection{Block Discarding attack}

\citet{bahack2013theoretical} introduced the Block Discarding attack, a concept similar to the selfish mining strategy proposed by \citet{eyal2018majority}. However, there are several key distinctions between these approaches. Firstly, they introduced a parameter called Network Superiority (\(ns\)), analogous to the communication capability of the attacker (\(\gamma\)) presented by \citet{eyal2018majority}. They analyzed the Block discarding attack through a hierarchical family of strategies denoted as \(st_{k}\) , where \(k\) represents different strategies \( (k = 0, 1, 2, \ldots) \) tailored to various combinations of attacker hash rates (\(p\)) and \(ns\) values. In this family, the selfish mining strategy proposed by \citet{eyal2018majority} is considered the simplest case, corresponding to the case when \(k=1\). Secondly, unlike the selfish mining strategy, which was initially designed for mining pools, \citet{bahack2013theoretical} argues that maintaining block secrecy within pools is highly challenging. Therefore, they propose that the Block discarding attack is more feasible for execution by solo miners rather than mining pools. This distinction emphasizes the tailored approach for solo miners, aiming to exploit the network’s vulnerabilities more effectively without the complexities involved in coordinating a mining pool's efforts. Another notable difference is that \citet{eyal2018majority} describes a process that continues indefinitely until the attacker's pool completely ousts all other miners. In contrast, \citet{bahack2013theoretical} argues that a more likely outcome is the establishment of a new equilibrium where there are fewer honest miners, but those remaining continue to maintain the same profit levels due to the difficulty adjustment making mining easier by lowering the difficulty level to maintain the same block rate. Alternatively, the study suggests a scenario where all honest miners might eventually leave the system entirely. 

\subsubsection{Difficulty Raising attack}

\citet{bahack2013theoretical} proposed another type of attack called the Difficulty Raising attack which aims to manipulate the Bitcoin difficulty adjustment mechanism to create a situation where an attacker’s chain can surpass the honest chain. This involves increasing the difficulty of the attacker’s chain in such a way that it eventually surpasses the difficulty of the honest chain, allowing the attacker’s chain to become the accepted chain. This is because in Bitcoin, the chain with the highest cumulative difficulty is considered the valid chain by the network as it follows the "longest chain" rule, which is essentially the chain with the most work put into it. In this attack, at the start of a new difficulty window, the attacker creates a secret chain of 2014 blocks, each with timestamps that are one second apart or the same as the first block's timestamp. The attacker ensures that the difficulty of these blocks is higher than what would be expected based on the honest chain. They do this by adjusting the difficulty setting for each block or set of blocks to be more challenging than the current difficulty level of the honest chain. As the attacker prepares for the the second window, they set the difficulty \textit{d} of this upcoming window to be significantly higher. This is done to make the total difficulty of the attacker’s chain surpass that of the honest chain. Once the attacker’s chain has a higher total difficulty, they reveal it to the network. According to Bitcoin’s consensus rules, the network will adopt the attacker’s chain as the valid chain, effectively discarding the honest chain.  

According to \citet{bahack2013theoretical}, while the probability of success is 1, the average time required for success is infinite. However, they show that the median time is finite and can be approximated as 
\[
\frac{\sqrt{e} \times 2015}{p}
\] units of 10 minutes where \( p \) is the ratio between the hash-powers of the attacker and the honest network.

\citet{bahack2013theoretical} further shows that as mining technology improves and hash power increases, both the honest network and the attacker’s hash power might grow. If both increase proportionally, the attack becomes easier, because the difficulty ratio between the attacker's chain and the honest chain becomes more favorable for the attacker. If hash power grows exponentially over time, the necessary difficulty ratio for the attack decreases, making it easier for the attacker to eventually succeed.

\subsubsection{Block Withholding attack}

The Block Withholding (BWH) attack is a mining attack in which a malicious miner participates in a mining pool while withholding valid blocks, thereby inflicting financial damage on the pool \citep{rosenfeld2011analysis}. The attacker initiates their attack by joining a mining pool just like any regular miner and contributing their computational power to the mining pool. The attacker starts to mine and submits Partial PoW solutions (PPoW) to the pool, which are required to demonstrate their participation and effort. If a miner successfully finds a Full Proof of Work (FPoW) solution that satisfies the network's difficulty requirements, they may choose not to submit this valid block to the pool operator. Instead, the attacker withholds the block, refraining from broadcasting it to both the network and the mining pool. By withholding the valid block, the mining pool misses out on the associated block reward and transaction fees. Consequently, this results in a reduction of the pool's overall earnings, which can lead to diminished profitability and may ultimately cause miners to leave the pool.

\subsection{Hybrid attacks analysis}
\label{sec:iv-b}

\begin{figure}[!t]
	\centering 
	\includegraphics[width=1.0\textwidth]{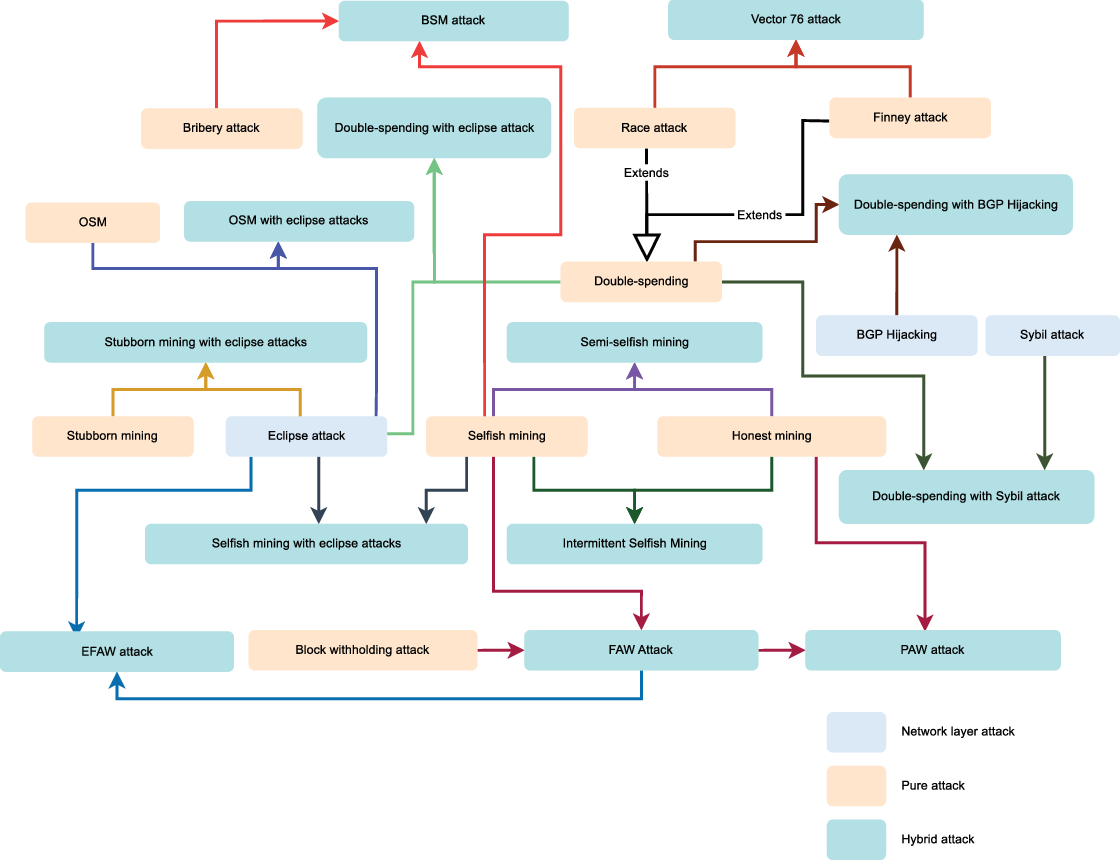}	
	\caption{Various Hybrid attack relationships} 
	\label{fig:hybrid_map}%
\end{figure}

In our study we define Hybrid attacks as sophisticated attack strategies that combine a pure attack with another pure attack or with another malicious or non-malicious strategy to potentially increase the success probability or the profitability of the attack. This section provides a comprehensive analysis on hybrid attacks. Figure \ref{fig:hybrid_map} illustrates the relationships and interactions between various attack strategies.

\subsubsection{Intermittent Selfish mining (ISM)} 
\citet{negy2020selfish} introduced a variant of the selfish mining strategy known as Intermittent Selfish Mining (ISM). In this approach, the attacker halts their selfish mining actions immediately following a difficulty adjustment. In particular, the attacker switches between selfish and honest mining strategy at every difficulty adjustment in Bitcoin. ISM strategy consists of two distinct phases. In the first phase, the  attacker engages in selfish mining with the intention of undermining the honest miners by withholding blocks and strategically releasing them to maximize their own block rewards. This phase mainly serves to invalidate the blocks mined by the honest participants; therefore, it sets the attacker in an advantageous position for the start of the next epoch. In the second phase, following the difficulty adjustment, the attacker transitions to honest mining. Given the reduced difficulty due to the slower block discovery rate induced by the earlier selfish mining phase, the mining process is more efficient. This lower difficulty results in a faster mining rate, benefiting all miners, but most significantly allowing the attacker to gain a higher relative reward in both phases. 

\subsubsection{Semi-selfish mining}

The selfish mining strategy can backfire if honest miners detect an abnormal forking rate, prompting them to abandon the blockchain and thereby reducing the revenue of selfish miners. Given that honest miners estimate network conditions, it is prudent for selfish miners to control the fork rate to maintain an acceptable level. Based on this idea, \citet{li2021semi} introduced a new form of selfish mining variant called semi-selfish mining, which is based on a Hidden Markov Decision Process (SMHMDP) \citep{ephraim2002hidden}. In this approach, miners masquerade as honest parties with a certain probability, aiming to reduce the forking rate and thus avoid detection. Unlike a traditional selfish miner, who consistently mines on their private chain—leading to a relatively higher forking rate by releasing blocks from the private chain to the public chain—a semi-selfish miner adopts a more nuanced strategy. A semi-selfish miner appends newly mined blocks to the public chain, thereby infiltrating a small portion of their computational power into the pool of honest miners. This strategy helps enhance the perceived power of honest miners because the blocks mined by semi-selfish miners are counted as honest blocks. By presenting themselves as honest parties, semi-selfish miners aim to maintain the forking rate at an acceptable level to avoid arousing suspicion and causing honest miners to leave the network. Despite this facade, attackers still manage to profit from selfish mining. By subtly influencing the blockchain and moderating the fork rate, semi-selfish miners can continue their exploitation while mitigating the risk of detection and the subsequent departure of honest miners.

\subsubsection{Selfish mining with Eclipse attacks}

\citet{heilman2015eclipse} showed that the success of the selfish mining strategy can be increased when it is combined with eclipse attacks. The effectiveness of selfish mining depends on the computational power of the adversary pool \((\alpha)\) and the communication capability of the adversary \((\gamma)\). As \(\gamma\) represents the computational power of the honest miners that end up mining on the adversary's chain during a fork, when \(\gamma\) is high, then \(\alpha\) can be small \citep{heilman2015eclipse}. A lower \(\alpha\) means that the pool needs to possess small computational power, making the attack easier. With an eclipse attack, the adversary can increase \(\gamma\) by isolating certain miners from the rest of the network. The isolated (eclipsed) miners are fed a false view of the blockchain that only shows the adversary's blocks. When eclipsed miners find blocks that would compete with the adversary’s chain, the adversary discards these blocks, preventing them from propagating through the network. By controlling what the eclipsed miners see, the adversary ensures that these miners continue to mine on the adversary’s chain. This effectively increases \(\gamma\) because the eclipsed miners unknowingly support the adversary’s chain rather than the honest chain. With a high \(\gamma\), even a small \(\alpha\) is enough for the adversary to successfully implement the selfish mining strategy.

\subsubsection{Stubborn mining with Eclipse attacks}

\citet{nayak2016stubborn} showed that an attacker can increase their revenue by combining their stubborn strategies with eclipse attacks. They presented two extremes of composing eclipse attacks with stubborn strategies, namely \textit{Collude} and \textit{Destroy}, and proposed a moderate yet effective strategy called \textit{Destroy if No Stake (DNS)}.

Under the \textit{Destroy} strategy, the attacker chooses to drop all blocks mined by the eclipsed victim. This particularly neutralizes or “destroys” the impact of the eclipsed victim's mining efforts, enabling the attacker to increase their revenue.

Under the \textit{Collude} strategy, the attacker enters into a collaborative agreement to mine together. Both the attacker and the eclipsed victim work on a single private chain, and the attacker agrees to accept the blocks mined by the eclipsed victim to extend this private chain. While the attacker and the eclipsed victim work together, the attacker still follows a stubborn mining strategy when communicating with the rest of the network.

Under the \textit{DNS} strategy, the attacker maintains a private blockchain while the eclipsed victim may mine on either the attacker’s chain or their own separate chain. The attacker uses specific criteria to decide whether to accept or reject blocks mined by the victim:

\begin{itemize}
    \item \textbf{Acceptance}: The attacker will accept a block from the victim only if it extends the attacker’s private blockchain. This means the block must build directly on the attacker’s chain.
    \item \textbf{Rejection}: The attacker will reject the victim’s block if:
    \begin{itemize}
        \item The attacker is not maintaining a private blockchain.
        \item The attacker and the victim are mining on separate chains.
    \end{itemize}
\end{itemize}

This selective approach enables the attacker to increase their own gain by employing a conditional strategy that combines elements of both \textit{Collude} and \textit{Destroy}. \citet{nayak2016stubborn} show that when \(\alpha=0\), an eclipse attacker can yield gains of up to 30\% by combining stubborn strategies compared to the naive utilization of eclipsed nodes.

\subsubsection{Double-spending with Sybil attacks}

\citet{iqbal2021exploring} showed how a double-spending attack can be combined with a Sybil attack. In this scenario, the attacker creates numerous fake identities, known as Sybil nodes, to interfere with regular communication and transaction processing in a blockchain network. The attacker initially generates several Sybil nodes, which appear as legitimate participants in the network. After that, the attacker initiates a transaction \( T_x \) and broadcasts it to the network. Honest nodes, which follow the standard protocol, verify this transaction and add it to their memory pool, where pending transactions wait to be included in the next block.

While the network processes \(T_x\), the attacker starts mining a private chain and simultaneously creates a double-spend transaction \(T_y\), which is included in the attacker’s private chain. The Sybil nodes are used by the attacker to intentionally delay the propagation of the new block containing \(T_x\). These nodes accomplish this by refraining from sharing information about the newly mined block with the rest of the network, effectively halting the propagation process. If honest nodes do not receive a new block within a certain time frame, they stop waiting and start mining the next block. The delay caused by Sybil nodes helps the attacker’s private chain catch up with or surpass the length of the honest chain.

If the attacker successfully mines enough blocks on their private chain, making it longer than the public chain, the network will accept the attacker’s private chain as the valid chain. When this occurs, the attacker’s double-spend transaction \(T_y\) becomes valid, and \(T_x\) is discarded. As a result, the attacker can reclaim the coins initially spent in \(T_x\) while also having successfully executed \(T_y\), effectively spending the same coins twice. Their study shows that it is not necessary for the attacker to control the majority of the network’s computing power; success can be achieved with just 32\% of the computational power when utilizing Sybil nodes to delay block propagation.

\subsubsection{Double-spending with Eclipse attacks}

\citet{iqbal2021exploring} demonstrated a type of hybrid attack that combines elements of an eclipse attack to execute a double-spending scheme, exploiting both 0-confirmation and \(N\)-confirmation mechanisms. In this attack, the adversary targets specific nodes within a blockchain network, rather than the entire network, as is typical with a Sybil attack. The attacker first identifies a target node, which could be a miner, a merchant, or any other relevant node. Subsequently, the attacker deploys an eclipse attack by flooding the victim node with connections from their own IP addresses, isolating it from the rest of the network. As a result, the victim node can only communicate with the attacker.

In the case of the 0-confirmation mechanism, the attacker sends a transaction \(T_x\) to the merchant's eclipsed node. The isolated merchant, believing the transaction is valid, proceeds to send the goods. Concurrently, the attacker initiates a second transaction \(T_y\) that double-spends the same amount of cryptocurrency and broadcasts this transaction to the rest of the network. Since the merchant’s node is eclipsed, it cannot broadcast \(T_x\) to the wider network. Consequently, when the network eventually confirms a transaction, it validates \(T_y\) and discards \(T_x\). This allows the attacker to successfully receive the goods without actually making a payment, as the legitimate transaction \(T_x\) never gets confirmed.

In the case of \(N\)-confirmations, where merchants may wait for a certain number of confirmations before releasing goods, the attacker isolates both the merchant and a fraction (\(n\)) of the miners responsible for confirming transactions. These miners are also eclipsed and only see the blockchain controlled by the attacker. Similarly, the attacker generates a transaction \(T_x\) for the merchant’s eclipsed node. The merchant waits for \(N-1\) confirmations, which are provided by the miners under the attacker's control during the eclipse. Believing that the transaction has been confirmed, the merchant releases the goods to the attacker. After receiving the goods, the attacker reconnects the eclipsed miners to the actual blockchain network. The blockchain view provided by the attacker becomes orphaned because it does not align with the actual network’s blockchain. As a result, the merchant's transaction \(T_x\) is never truly confirmed in the real blockchain, enabling the attacker to acquire goods without payment.

As highlighted by \citet{iqbal2021exploring}, unlike Sybil attacks, which affect the entire network, eclipse attacks focus on specific nodes. This specificity makes them more covert and harder to detect. In contrast to 51\% attacks, which require controlling the majority of the network's computational power, eclipse attacks can be executed with fewer resources by targeting specific nodes. Furthermore, their study emphasizes that eclipse-based double-spending is particularly dangerous in systems that utilize 0-confirmations or \(N\)-confirmations for faster transactions, as it exploits the trust merchants place in unconfirmed or minimally confirmed transactions.

\subsubsection{Double-spending with BGP Hijacking}

The Border Gateway Protocol (BGP) is a protocol used to exchange routing information between different networks, such as Internet Service Providers (ISPs) or Autonomous Systems \citep{iqbal2021exploring, quoitin2005modeling}. These networks are independently operated and collectively comprise the broader internet infrastructure. 

BGP hijacking is a type of routing attack in which an attacker manipulates the BGP protocol to make false routing announcements, diverting traffic from its intended path to a path they control \citep{iqbal2021exploring}. In this attack, the adversary falsely claims ownership of specific IP address ranges (prefixes), causing traffic destined for those addresses to be rerouted to the attacker’s network. This manipulation can lead to network partitioning, including the separation of blockchain data, across the internet. 

By hijacking IP prefixes, the attacker can effectively isolate significant portions of the blockchain network. For instance, they can isolate up to 50\% of the Bitcoin network's hash rate by hijacking fewer than 100 BGP IP prefixes. This isolation allows the attacker to slow down the propagation of new blocks across the network. With delayed block propagation, an attacker can execute either 0-confirmation or \(N\)-confirmation double-spending attacks by exploiting the time lag between when a transaction is broadcast and when it is confirmed on the blockchain.

\subsubsection{Optimal selfish mining with Eclipse attacks}

\citet{gervais2016security} introduced a more comprehensive and realistic mining MDP model for optimal selfish mining, capturing various real-world blockchain parameters such as stale block rates, mining power, mining costs, and eclipse attacks. They modeled the mining problem as a single-player decision problem. The state space \(S\) was defined as a four-tuple in the form \( (l^{(a)}, l^{(h)}, b_{e}, fork) \), where \(b_{e}\) represents the blocks mined by the eclipsed victim. 

To solve this single-player decision problem and determine the optimal selfish mining policy, they employed a method similar to that proposed by \citet{sapirshtein2017optimal}. Their examination focused on how stale block rates influence an adversary's relative gain. The stale block rate parameter effectively captures various blockchain characteristics, such as block sizes, intervals, network delays, propagation mechanisms, and overall network configuration, thereby enhancing the realism of the mining MDP model. 

The findings indicate that as the adversary's hash rate increases, their relative revenue surpasses the upper bound of mining gains established by \citet{sapirshtein2017optimal}. Additionally, their work explored the impact of eclipse attacks on selfish mining, demonstrating that as the adversary’s hash rate grows, their capabilities for selfish mining significantly strengthen.

\subsubsection{Fork after Withholding attack}

\citet{kwon2017selfish} introduced a new attack known as the Fork after Withholding (FAW) attack. This approach combines features of the Block Withholding attack with strategies from selfish mining, specifically targeting the consensus mechanism within blockchain networks. The hybrid nature of this attack allows an attacker to achieve higher profits than a standard BWH attack, regardless of their computational power or network capability.

In a FAW attack, the attacker joins a target mining pool and strategically divides their computational power between honest mining and infiltration mining, akin to a BWH attack. In a traditional BWH attack, when the attacker discovers a valid FPoW solution, they withhold it without submitting it to the pool manager. In contrast, during a FAW attack, the attacker refrains from immediately propagating the FPoW solution to the pool manager. Instead, they wait for an external honest miner to publish their FPoW solution first. Once this occurs, the attacker then propagates their withheld FPoW solution to the pool manager. 

If the pool manager accepts the submitted FPoW from the attacker, it propagates it to the network, resulting in the creation of a fork. All participants in the Bitcoin network must then choose between the competing branches. If the attacker’s block is selected as the valid chain, the target pool receives the reward, and the attacker is also compensated by the pool. By employing this method, a FAW attacker can secure additional rewards regardless of the specific outcome, as the successful inclusion of the attacker’s block by the target pool ensures that they benefit. 

This makes the FAW attack particularly potent and profitable compared to a BWH attack, as it maximizes the attacker’s gains through strategic block propagation. The authors demonstrate in their study that a FAW attacker can earn significantly higher rewards—ranging from one to four times more—than a BWH attacker within a large pool that controls approximately 20\% of the computational power of the entire Bitcoin network. Furthermore, the study extends the FAW attack to multiple pools, enabling the attacker to accumulate even greater rewards. Their analysis reveals that if an attacker targets four popular mining pools with the FAW attack, their additional reward can be approximately 56\% greater than that of a BWH attacker.

\subsubsection{Fork after Withholding attack with Eclipse attacks}

\citet{wang2023efaw} introduced a new attack model that integrates the characteristics of a Fork after Withholding attack with eclipse attacks. This hybrid approach, referred to as the Eclipsed Fork After Withholding (EFAW) attack, proves to be more profitable than standard FAW attacks. In an EFAW attack, the attacker divides their computational resources between honest mining and infiltration mining. Subsequently, they initiate an eclipse attack on nodes within the target mining pool. By controlling the communication of these nodes, the attacker can manipulate which Proof of Work (PoW) results are submitted and accepted by the network.

When a node in the victim pool discovers a PPoW solution, the attacker may choose to withhold the block, thereby excluding it from the consensus process and preventing it from being accepted by the network. This increases the attacker’s share of mining rewards, as the victim’s work goes unrewarded. For a FPoW solution, the attacker can strategically release withheld blocks to create forks in the blockchain, similar to the original FAW attack. By timing this correctly, the attacker can create branch points in the blockchain that further enhance their rewards. Their study analyzed the effectiveness of the EFAW attack against one and two victim pools.

For an EFAW attack targeting a single victim pool, let us consider two pools: the attacker's mining pool \(P_1\) and the victim pool \(P_2\). The computational power is allocated such that part of \(P_1\)'s computational power is dedicated to honest mining within its own pool, while the remaining computational power is used to infiltrate \(P_2\). The attacker then takes advantage of the eclipse attack to control certain nodes within \(P_2\), referred to as eclipsed miners. These miners are isolated from the broader network and only receive and send information that the attacker controls. The manager of \(P_1\) regulates the proof submissions in the following manner.

\begin{itemize}
    \item Honest Miners in Pool A: Their FPoWs are published immediately.
    \item Eclipsed Miners in Pool B: Their PPoWs are discarded by \(P_1\)'s manager.
    \item Infiltrator Miners in Pool A: Their FPoWs and PPoWs are retained by \(P_1\)'s manager.
\end{itemize}

The execution of the attack can be analyzed with following cases.

\begin{enumerate}
    \item  When no FPoW is found: If neither the infiltration miners nor the eclipsed miners find an FPoW, but an honest miner in \(P_1\) finds one, it is immediately submitted.

    \item When FPoWs are found:
    \begin{itemize}
        \item If an FPoW is found by an honest miner in \(P_1\), any FPoWs from the infiltrators or eclipsed miners are discarded by the manager.
        \item If an FPoW is found by an honest miner in \(P_2\), the manager in \(P_1\) will also discard the FPoWs from infiltrators or eclipsed miners.
        \item If another honest miner in \(P_1\) submits a valid block, it results in a fork in the main chain. The retained FPoWs contribute to this fork, disrupting the network consensus and potentially benefiting \(P_1\).
    \end{itemize}
\end{enumerate}
    
\citet{wang2023efaw} developed a reward formula for the victim pool, demonstrating mathematically that the rewards from the EFAW attack are equal to or exceed those of FAW attacks. Additionally, they showed that if the eclipse attack fails to isolate the miner nodes of the victim pool, the EFAW attack essentially becomes an FAW attack. Furthermore, their study analyzed the scenario when the attacker targets not just one but two honest mining pools simultaneously. By extending the previous example, consider a second victim pool \(P_3\). The attack involves both infiltration mining and eclipse attacks on these victim pools. Hence, the infiltration computation power is allocated to mine and infiltrate both \(P_2\) and \(P_3\). Now \(P_1\) launches eclipse attacks on \(P_2\) and \(P_3\). This allows \(P_1\) to discard any PPoWs submitted by the eclipsed miners and to retain FPoWs submitted by infiltrator miners. When an infiltrator miner or an eclipsed miner finds an FPoW, \(P_1\)'s manager retains it. Whenever other honest miners send their blocks, the manager of \(P_1\) immediately submits the withheld FPoW, resulting in either a two-branch or three-branch fork in the blockchain.

\subsubsection{Power Adjusting Withholding}

In a Fork after Withholding attack, the attacker divides their hash power between honest mining and infiltration mining. The honest mining portion yields the full reward for mining the block, while the infiltration mining portion provides only the attacker's fair share of the pool’s revenue. \citet{gao2019power} argue that when the mining pool is relatively large and the attacker’s chain has a lower chance of being chosen as the main chain, allocating more power to honest mining becomes more profitable. This is because the profit from infiltration mining diminishes once a FPoW is found, resulting in wasted mining power on less attractive rewards. Based on this observation, \citet{gao2019power} proposed the Power Adjusting Withholding (PAW) attack, which eliminates the static power-splitting strategy in the FAW attack by dynamically adjusting mining power between honest and infiltration mining. This dynamic adjustment allows the attacker to increase their revenue by allocating more power to the more attractive reward. Their analysis demonstrates that PAW attacks consistently enable the attacker to earn a higher reward, potentially increasing their earnings up to 2.5 times more compared to FAW attacks. Furthermore, they show that PAW attacks can avoid the "miner's dilemma," where the larger pool typically prevails when multiple mining pools engage in PAW attacks against each other.

\subsubsection{Bribery Selfish Mining}
The Bribery Selfish Mining (BSM) attack integrates bribery racing strategy with selfish mining attacks \citep{gao2019power}. In selfish mining, stubborn mining, and FAW attacks, when a fork occurs in the blockchain, the attacker continues to mine on their previously private chain. In contrast, honest miners make their choice based on the dissemination of notifications, deciding to mine on either branch. The attacker can increase the chances of their chain being recognized as the longest by encouraging more honest miners to contribute to it. Bribery attacks offer a particularly insidious method for achieving this goal. By offering financial incentives (bribes), the attacker can increase the probability that their branch is selected as the winning chain. By bribing other miners to mine on their branch, the attacker improves their chances of winning the competition to extend the blockchain. Consequently, other miners, motivated by the prospect of a higher reward, will choose to extend the branch published by the attacker. However, as outlined by \citet{gao2019power}, plain bribery attacks do not inherently yield any gain for the attacker. Therefore, bribery attacks are often combined with other strategies. The BSM attack incorporates bribery transactions into the attacker's branch, providing rewards that any miner can claim by choosing to mine on that branch. This strategy aims to lure other miners to join the attacker's side. Their findings indicate that BSM can yield an additional 10\% reward for the attacker compared to standard selfish mining. However, their analysis also indicates that BSM attacks suffer from the "venal miner's dilemma". In this scenario, all other miners choose to extend the attacker's branch in pursuit of higher rewards, but this collective behavior ultimately results in losses that are worse than those incurred through honest mining.

\subsubsection{Vector76 attack}
The Vector76 attack, named after a user on the Bitcointalk forums is a variant of double-spending attack that combines elements of the Race attack and the Finney attack \citep{vector76_attack,vector76_attack_2}. This attack involves manipulating the network by targeting a small subset of nodes and taking advantage the delay in transaction confirmation \citep{vector76_attack_2}. This attack involves the following key components.
\begin{itemize}
    \item An \textit{attacker} is a malicious miner who aims to exploit the transaction confirmation process.
    
    \item An \textit{Honest Miner} who follows the standard protocol and is not involved in the attack.
    
    \item A well-connected, high-volume node that broadcasts transactions widely (say \textit{Node A}).
    
    \item A node that quickly verifies and broadcasts transactions from Node A (say \textit{Node B}).
    
\end{itemize}

The attack involves the following steps.
\begin{enumerate}
    \item The attacker and the honest miner both aim to solve a block at the same time. Each miner includes a transaction in their block. The goal is for the attacker to use their block to double-spend by exploiting the timing of broadcast and confirmation.
    \item The attacker targets \textit{Node A} and \textit{Node B} and submits their block to Node A at the same time the honest miner submits theirs. 
    \item Node A picks up and broadcasts the transaction, which is also quickly verified by \textit{Node B} and other nodes connected to \textit{Node A}.
    \item The attacker requests a withdrawal from \textit{Node A}, who processes the transaction and provides the deposit. Since \textit{Node A} and \textit{Node B} are quick in verifying, the transaction appears to be valid to the recipient.
    \item If the attacker's block is accepted and extended by other miners, they receive the block reward, making the double-spending successful. The attacker effectively gets the block reward and keeps the double-spent funds. However, if the attacker fails then  the honest miner’s block is ultimately accepted and the attacker's block becomes stale or orphaned, the initial deposit transaction is still valid. The attacker retains the withdrawn funds but does not receive the block reward.
\end{enumerate}

\subsection{Multiple pool attack analysis}
\label{sec:iv-c}

This section examines how competing mining pools deploy selfish mining attacks and Block Withholding attacks against each other. By analyzing interactions between pools, this analysis provides insights into the dynamics of these attacks when executed by different mining pools against each other. The goal is to understand how this competition affects the overall profitability and the distribution of rewards across pools. 

\subsubsection{Selfish mining in multiple pools}

Various studies have employed models to analyze the selfish mining strategy, typically assuming the presence of a single selfish miner while acknowledging the potential for multiple colluding pools nearing the profitability threshold \citep{eyal2018majority,nayak2016stubborn,sapirshtein2017optimal}. However, in a realistic setting, multiple mining pools with significant hash power could engage in selfish mining simultaneously.

\citet{liu2018strategy} introduced a new model of a PoW-based blockchain that allows for the presence of multiple independent attackers. In this context, "independent attackers" refers to attackers whose decision-making processes are independent of one another, although their state transitions are influenced by other miners. According to their formulated model, each attacker encounters a single-player decision problem. Given that each attacker must maintain their state, they defined the attacker's state as a 3-tuple \( T = (lead, f_{1}, f_{2}) \), where \( lead \) indicates the attacker's lead over the honest chain, \( f_{1} \) denotes whether there is a fork in the main chain (indicating the existence of competition), and \( f_{2} \) indicates if the attacker is involved in this competition. The action space for each attacker includes \( Hold \), \( Match \), \( Override \), \( Adopt \), and \( Publish \). This action space is similar to the previously discussed MDP models, but with an additional action, \( Publish \), which pertains to the event where \( attacker_{i} \) publishes the head of their blocks. To evaluate the performance of their mining strategy for each attacker, they adopted the relative stale block rate associated with each attacker.

Based on this model, \citet{liu2018strategy} proposed a new strategy called publish-n (\( P_{n} \)). The intuition behind this strategy is to enable the attacker to shorten their private chain, which is advantageous in situations where they hold a long private chain but still lag behind the main chain. The value \( n \) acts as a trigger; when the attacker's state reaches \( n \), they will either publish the first block of their private chain or execute the \( Override \) action, depending on whether they have found the next block. \citet{liu2018strategy} simulated selfish mining, stubborn mining, and their publish-n strategy in a PoW-based blockchain environment with multiple attackers. Their results indicate that publish-n can surpass selfish mining by an efficiency of up to 26.3\%.

\citet{bai2019deep} introduced a novel MDP with a finite number of states to depict the state transitions between public and private chains, considering the presence of two selfish miners within the Bitcoin network. The selfish mining scenario was modeled by including an honest pool representing all legitimate miners within the network, alongside two independent selfish miners who are unaware of each other's non-compliant behavior. In their paper, the authors addressed the question of profitability by considering both the hash rates of the attackers and the adjustments made to the mining difficulty. They showed that when the hash rates of selfish miners are maintained at 22\%, the attackers can start benefiting from selfish mining after 51 rounds of mining difficulty adjustments, which translates to about 714 days in Bitcoin. However, if the hash rates are increased to 33\%, this timeframe decreases significantly to just 5 rounds, or approximately 70 days.

In contrast to the 3-player game utilizing a finite state MDP model as proposed by \citet{bai2019deep}, \citet{zhang2020analysing} conducted simulations of attacks with an infinite range of states and a larger number of players. For their analysis, they extended the model from their previous work \citep{zhang2020simulation} to evaluate scenarios involving multiple independent selfish miners. This extension allowed them to determine the mining power thresholds at which selfish mining becomes advantageous for all attackers. Specifically, to analyze the collective benefit for all attackers, they introduced the concept of the Common Beneficial Area (CBA). CBA represents the range of mining powers across different attackers such that all attackers simultaneously profit from selfish mining. Their findings indicate that the threshold at which selfish mining becomes beneficial decreases as the number of players increases. Notably, in a 5-player game (comprising 4 attackers and 1 honest miner), the beneficial threshold drops to 15\%, and it further decreases to 12\% in an 8-player game. However, the study argues that sustaining a multi-player selfish mining attack with more than 7 attackers is improbable in realistic scenarios, as the CBA diminishes with an increasing number of players. In fact, the CBA vanishes when the number of players exceeds 8, rendering the attack unsustainable. Nevertheless, the attack becomes feasible when the number of players surpasses this threshold.

\citet{bai2023blockchain} conducted an analysis of the profitability associated with selfish mining involving multiple attackers within a blockchain network. They formulated a Markov chain model to calculate the relative revenue of each attacker in a system with multiple selfish mining participants. However, as highlighted in their study, in such an environment, attackers cannot observe each other's private chains. Additionally, due to the anonymity inherent in blockchain technology, blocks released by the honest miner and the other attackers on the public chain cannot be distinguished. Based on this observation, \citet{bai2023blockchain} proposed a novel mining strategy for miners operating under incomplete information. This strategy is grounded in a family of Partially Observable Markov Decision Processes (POMDP), which features a large state space. It defines a strategic mining approach that offers greater rewards compared to both standard selfish mining and honest mining. To calculate the near-optimal mining policy, they employed AEMS2 \citep{ross2007aems}. Their analysis focused on the interactions among three types of chains: the strategic attacker, the basic selfish mining attacker, and the honest miner. The results indicate that the profit threshold for the strategic attacker decreases significantly, from 29.44\% to approximately 2\%, when the basic selfish mining attacker possesses a 34\% hash rate. This novel approach allows the strategic attacker to earn higher rewards in an environment with multiple selfish miners, showcasing the effectiveness of the POMDP-based strategy.

\subsubsection{Block Withholding attack in multiple mining pools}

Numerous studies have examined various methods of block withholding attacks, including those targeting dual mining pools \citep{li2020mining,wu2019equilibrium,qin2020optimal,yang2019game,hu2019game,wang2019pool}, multiple mining pools \citep{haghighat2019block,kim2019mining,wang2019research}, and hybrid block withholding attacks \citep{ke2019ibwh,chang2019uncle,wang2020optimal}. These studies provide a comprehensive understanding of the mechanisms and impacts of such attacks on the blockchain ecosystem.

A Block Withholding attack in a dual mining pool is a specific form of BWH attack that combines elements of block withholding and exploits the dynamics of mining pools. In this scenario, two mining pools engage in launching BWH attacks against each other \citep{li2020mining}. In this attack, the attacker chooses to withhold blocks from one pool while submitting them to the other, reducing the profitability of competing pools. A mining pool may exploit this strategy to enhance its competitive advantage by deploying miners to infiltrate rival pools or disrupt their operations. By executing Block Withholding attacks against these competing pools, the mining pool can generate additional revenue, ultimately increasing the profitability of its own operations \citep{li2020mining}. For instance, if a pool \(P_{1}\) wants to infiltrate a potential competitive pool \(P_{2}\), \(P_{1}\) sends a malicious miner \(M\) to \(P_{2}\) making \(M\) to submit partial PoW solutions to \(P_{2}\) and if \(M\) finds a valid block that meets the network’s difficulty, instead of submitting the valid block to \(P_{2}\), \(M\) withholds it and does not broadcast it to the network. Therefore, without effectively mining, yet still receiving rewards which are then redirected back to the original mining pool \(P_{1}\), thereby increasing the income of mining pool \(P_{1}\) at the expense of mining pool \(P_{2}\). Sometimes, the miners who were originally supposed to infiltrate pool \(P_{2}\)  may betray pool \(P_{1}\) by honestly mining on \(P_{2}\) which in turn reducing the revenue of \(P_{1}\). Moreover, when two pools launch BWH attacks on each other, it can lead to a version of iterative prisoner’s dilemma called “miner’s dilemma”. It means that both of them will suffer from a loss under the Nash equilibrium \citep{eyal2015miner}. For instance, if  \(P_{1}\) chooses to attack on  \(P_{2}\), it will result in a loss in its revenue, and can retaliate by attacking and increasing its revenue. However, when both  \(P_{1}\) and \(P_{2}\) attack, both earn less than they would have if neither attacked at Nash equilibrium. \citet{li2020mining} proposed a game theory \citep{fudenberg1991game} based multi-pool mining model for mining pools operating under the PoW consensus algorithm. This model incorporates a reward and punishment system designed to capture the attack behaviors of mining pools in a blockchain network. In their model, when a mining pool decides not to engage in attacks, it receives an additional reward, denoted as \( 0 \leq a \leq 1 \), provided by the system. Conversely, a mining pool that opts to attack is subjected to a penalty of \( k \cdot a \) (where \( k \geq 1 \) is the penalty-to-reward ratio). They examined the Nash equilibriums associated with both pure strategies—where a mining pool consistently employs a particular strategy—and mixed strategies—where the mining pool randomly chooses a strategy based on certain probabilities. Additionally, they developed a game model focused on mining pools in the context of BWH attacks. This model took into account two key factors: the infiltrate rate, which represents the percentage of malicious miners dispatched by the attacking pool to infiltrate the victim pool, and the betrayal rate, which indicates the proportion of malicious miners who choose to honestly mine in the victim pool rather than supporting the attacking pool. Their analysis aimed to assess the Nash equilibrium as well as the implications of the infiltrate rate within that equilibrium. The payoff matrix for two mining pools, \(P_{1}\) and \(P_{2}\), representing their choices between not attacking (\(N\)) and attacking (\(A\)), is presented in Table \ref{tab:payoff_1}. \(p_{1},p_{2}\) denotes the gains by the pools \(P_{1}\) and \(P_{2}\) respectively. \(a\) refers to the additional reward provided by the system, while \(k\) denotes the ratio of penalties incurred in relation to the rewards when a pool decides to launch an attack. Additionally, \( d' \) signifies the average profit for a pool that has decided to pursue an attack strategy, and \( d' \) represents the overall average benefit derived from competing mining pools when that same pool chooses to engage in an attack. For instance, if both \(P_{1},P_{2}\) decide not to attack \(P_{1}\) receives a payoff of \(p_{1}\) and \(P_{2}\) receives a payoff of \(p_{2}\).

\begin{table}[h]
    \centering
    \begin{tabular}{p{1.5cm} p{3cm} p{3cm}}
        \toprule
        \diagbox{\(P_{1}\)}{\(P_{2}\)} & \(N\) & \(A\) \\
        \midrule
        \(N\) & \(p_{1},p_{2}\) & \(p_{1}+a-d,p_{2}-k \cdot a+d\) \\
        \(A\) & \(p_{1}-k \cdot a + d,p_{2}+a-d\) & \(p_{1}-k \cdot a - d',p_{2}-k \cdot a +d'\) \\
        \bottomrule
    \end{tabular}
    \caption{Payoff matrix in the model proposed by \citet{li2020mining}}
    \label{tab:payoff_1}
\end{table}

\citet{wu2019equilibrium} introduced a model that allows two individuals to either collaborate with one another or to implement a Block Withholding strategy within a mining pool. In contrast to the model presented by \citet{li2020mining}, this model incorporates two additional costs for a mining pool: the cost of cooperation and the cost of generating partial proofs of work. They extended their model to calculate equilibrium by associating these costs directly to the computational power of the pool. For example, increasing the computational power of the pool results in greater cost when generating the PoW solution. Furthermore, they introduced a parameter called "payoff per time" (\(f(a)\) as a function of the total computational power (\(a\)) of all pools. This parameter effectively captures the total revenue of a pool given the total computational power of all pools when calculating the Nash equilibrium. The payoff matrix in the extended model which assumes the resource consumption is related to computational power for two mining pools \(P_{1}\),\(P_{2}\) is presented in Table \ref{tab:payoff_2}. \(C\) indicates choosing honest mining, \(A\) indicates choosing BWH attack, \(a_{1},a_{2}\) represent the computational powers of pools \(P_{1}\),\(P_{2}\), \(f()\) denotes the payoff of per time as a function of total computational power, \(C(a_{1})\),\(C(a_{2})\) denote the computational powers per time consumed for honest mining by \(P_{1}\),\(P_{2}\) and \(C_{p}(a_{1})\),\(C_{p}(a_{2})\) denote the costs per time for PPoW by pools \(P_{1}\),\(P_{2}\) respectively.

\begin{table}[h]
    \centering
    \begin{tabular}{p{1.5cm} p{4.5cm} p{4.5cm}}
        \toprule
        \diagbox{\(P_{1}\)}{\(P_{2}\)} & \(C\) & \(A\) \\
        \midrule
        \(C\) & 
        \begin{tabular}{@{}c@{}}
            \( f\left(a_1 + a_2\right) \cdot \frac{a_1}{a_1 + a_2} - C\left(a_1\right) \), \\
            \( f\left(a_1 + a_2\right) \cdot \frac{a_2}{a_1 + a_2} - C\left(a_2\right) \)
        \end{tabular} & 
        \begin{tabular}{@{}c@{}}
            \( f\left(a_1\right) \cdot \frac{a_1}{a_1 + a_2} - C\left(a_1\right) \), \\
            \( f\left(a_1\right) \cdot \frac{a_2}{a_1 + a_2} - C_p\left(a_2\right) \)
        \end{tabular} \\
        \midrule
        \(A\) & 
        \begin{tabular}{@{}c@{}}
            \( f\left(a_2\right) \cdot \frac{a_1}{a_1 + a_2} - C_p\left(a_1\right) \), \\
            \( f\left(a_2\right) \cdot \frac{a_2}{a_1 + a_2} - C\left(a_2\right) \)
        \end{tabular} & 
        \begin{tabular}{@{}c@{}}
            \( -C_p\left(a_1\right) \), \\
            \( -C_p\left(a_2\right) \)
        \end{tabular} \\
        \bottomrule
    \end{tabular}
    \caption{Payoff matrix in the model proposed by \citet{wu2019equilibrium}}
    \label{tab:payoff_2}
\end{table}

\citet{qin2020optimal} proposed an optimal BWH attacking model for two pools \(P_{1}\),\(P_{2}\) where one \(P_{1}\)  can attack the \(P_{2}\) but \(P_{2}\) cannot attack \(P_{1}\). Different from previous studies, they incorporated the probability of generating complete PoW solution by each pool in order to formulate their model. In their study, they considered two mining pools \(P_{1}\),\(P_{2}\) with computational powers \(c_{1}\) and  \(c_{2}\) such that  \(P_{1}\) attacks  \(P_{2}\) by utilizing  \(y_{1,2}\) of its computational power of  \(c_{1}\) but  \(P_{2}\) cannot attack  \(P_{1}\) as it performs honest mining. They formulated that the pool \(P_{1}\) can maximize its revenue by finding an optimal computational power to attack  \(P_{2}\). Furthermore, they analyzed the conditions of the influence of determining complete solution probabilities in pools for attacking to be profitable.

Previous studies have predominantly analyzed BWH attacks in the context of interactions between two mining pools. However, in the real world, multiple mining pools exist, and these pools can launch BWH attacks against one another. As the number of mining pools and miners increases, directly applying dual mining pool methods for optimizing block withholding attacks becomes impractical. Most researchers have focused on the simplistic scenario of a one-shot game between only two mining pools attacking each other. In reality, the competitive landscape involves multiple pools of varying sizes. \citet{haghighat2019block} addressed this complexity by formulating the BWH attack as a stochastic game with finitely many states and actions. They introduced the concept of dynamic migration of miners among mining pools. When miners decide to migrate from one pool to another, for instance because of their average revenue reduces, their destination is not predetermined, as they lack information about which pools are currently being attacked or are likely to be attacked in the future. Consequently, migrating miners make stochastic choices regarding their new destination, adding a layer of unpredictability to the game. This approach provides a more realistic framework for understanding and analyzing BWH attacks in a multi-pool environment. In their game model, the authors considered a total number of \(S_{T}\) miners, distributed across \(n\) mining pools of varying sizes as well as solo mining. Each pool has an associated parameter called attractiveness \( A \in \mathbb{R} \), which determines the willingness of miners to join or remain in the pool. The attractiveness parameter ranges from \(A=0\), indicating the minimum possible attractiveness, to \(A=1\), indicating the maximum possible attractiveness. When a pool is subjected to a BWH attack, its revenue decreases, thereby reducing its attractiveness. Miners perceive that this pool is less fortunate or is being targeted by other pools, leading to a decrease in its attractiveness. The attractiveness of a pool directly influences its size, as miners decide at the end of each round whether to remain in their current pool or migrate to another based on the attractiveness levels. To analyze the dynamics of this game, the authors incorporated Tile Coding RL algorithm \citep{bowling2002scalable}. Since the ultimate goal of the pool is to increase its income by increasing its size i.e. the total number of miners in the pool, to evaluate the game they modeled the utility of each pool in each round proportional to the change in its size (Equation \ref{eq:utility_1})
\begin{equation} \label{eq:utility_1}
u_i^t = \frac{S_i^t - S_i^{t-1}}{S_T}
\end{equation}
where \(S_i^t\),\(S_i^{t-1}\) denotes the size of the pool \(i\) in the rounds \(t\) and \(t-1\) respectively.

\section{Design guidelines}
\label{sec:v}

Decentralized consensus mechanisms are the foundation of blockchain technology, playing a crucial role in cryptocurrencies like Bitcoin. Mining, as a key component of these mechanisms, is vital for maintaining the stability and integrity of the blockchain. However, as the adoption and usage of cryptocurrencies continue to grow, attacks on consensus and incentive mechanisms pose significant threats to the stability and integrity of these networks.

Hence, in this section, we provide a set of guidelines that future research must focus on to address and mitigate these risks: 

\begin{itemize}
    \item Future research should focus on designing and implementing more robust consensus protocols that are inherently resistant to attacks like selfish mining and its variants. Variants such as semi-selfish mining provide attackers with additional advantages by allowing them to appear as honest miners, increasing the likelihood of a successful attack. Furthermore, optimal selfish mining with reinforcement learning enables attackers to dynamically learn the optimal selfish mining strategy without prior knowledge of blockchain parameters and adapt their mining policies in environments where these parameters are constantly changing. Therefore, the potential and effectiveness of these strategies should be further studied to develop robust countermeasures. This could involve creating protocols or modifying the existing protocols to detect abnormal forking rates in the main branch to minimize the benefits gained by attackers.
    
    \item Attackers can increase their profitability by combining pure strategies with network layer attacks, such as eclipse attacks and BGP Hijacking. For example, combining selfish mining with eclipse attacks allows an attacker to increase their revenue by isolating the victim’s mining power to advance their private chain. Therefore, future work should focus on improving the security and redundancy of communication channels within blockchain networks. This could involve developing decentralized, fault-tolerant routing protocols that make it more difficult for attackers to isolate nodes or manipulate network traffic.

    \item Incentive mechanisms should be redesigned to discourage behaviors that lead to attacks such as pool hopping and block withholding, which can financially harm mining pools. Pool Hopping attacks involve attackers frequently switching between mining pools to maximize their financial gain. This behavior can make the pool’s revenue distribution less efficient and result in higher operational costs, including expenses for server maintenance, bandwidth, and infrastructure, without receiving proportional rewards. As a result, the pool may spend resources on mining without capturing rewards as frequently, reducing overall efficiency. Block withholding attacks further reduce a pool’s gains by causing attackers to withhold FPoW solutions that meet the network's target difficulty. Future research should focus on redesigning incentive mechanisms that reward honest mining and penalize such adversarial behaviors, thereby protecting the integrity and stability of mining pools.  

    \item To prevent attacks like Difficulty Raising, future work should aim to improve the security of difficulty adjustment algorithms within blockchain networks. The focus should be on creating algorithms that respond more dynamically to sudden changes in network hash power, making it harder for attackers to manipulate the difficulty level. Because the difficulty adjustment mechanism is designed to adjust the difficulty level based on the overall network hash power and block discovery times. If the attacker is manipulating difficulty, the honest network might not adjust its difficulty quickly enough to keep up with the changes. The attacker exploits this delay or inefficiency in the difficulty adjustment to make their chain’s difficulty higher, giving their chain an advantage over the honest chain.

     \item To effectively detect or mitigate double-spending attacks, future research should focus on several key strategies. Enhancing transaction confirmation methods is essential; this might include improving multi-confirmation protocols that require multiple confirmations from different nodes or over extended time periods, as well as cross-chain verification to ensure transactions are not spent more than once across different networks. Real-time monitoring systems is another key aspect to detect such attacks. These systems should be developed to track transactions and detect anomalies or suspicious activities immediately, using advanced behavioral analysis to identify potential double-spending attempts. Sophisticated detection algorithms and fraud detection systems that utilize machine learning or statistical methods can improve the ability to spot double-spending attempts. Additionally, improving consensus mechanisms and ensuring better network coordination can strengthen the blockchain's resilience against such attacks. By addressing these areas, future research can significantly improve blockchain security and reduce the risk of successful double-spending attacks.

     \item Future research should aim to make bribery based attacks, like Bribery Selfish Mining harder and more expensive to implement. One way to do this is to put in place tools and systems that increase transparency in mining process and transactions. These tools can help to monitor miner behavior more closely and identify patterns that indicate potential bribery thus making it harder for attackers to secretly influence other miners.
     It is also essential to design ways that make it tougher to offer hidden money rewards. This could involve making the flow of transactions more visible and setting up strong checks on money transfers linked to mining. By exploring following strategies, future research can reduce the effectiveness bribery attacks and ensure a safer and fairer mining environment.

     \item To detect and prevent attacks that rely on mining centralization, such as 51\% attacks, future research should focus on strategies to decentralize mining power more effectively. This involves developing and promoting methods to distribute hash power across a wider range of participants, particularly among smaller, independent miners.

     \item Special attention should be given to attacks that exploit forking of the main branch. Forking-based strategies such as selfish mining and its variants, FAW attacks, and BR attacks take advantage of these forks to execute their adversarial strategies. Future research should concentrate on developing efficient methods for detecting and recovering from such forks. This could involve designing advanced algorithms capable of identifying abnormal or suspicious forks in the blockchain. Moreover, these algorithms should ensure that the blockchain quickly converges on the valid chain, to gain a quick recovery from such attacks. Research should also explore integrating real-time monitoring systems that can detect and address forks as they occur, thus enhancing the resilience and reliability of the blockchain against forking-based attacks.

\end{itemize}

\section{Conclusion}
\label{sec:vi}

In conclusion, this study has provided a comprehensive examination of various attacks on consensus and incentive mechanisms on PoW- based blockchain networks by exploring attacks categorized under pure attacks, hybrid attacks and evaluating selfish mining and block withholding attacks in multiple pool environments. Future work should focus on the development of robust detection and recovery methods for different fork-based attacks, improving miner incentive structures, strengthening network communication layers, and enhancing transparency to mitigate the risks posed by these attacks. By focusing on these areas, future research can contribute to more secure and reliable blockchain systems, ensuring that blockchain technology remains robust against evolving threats.

\bibliographystyle{elsarticle-num-names}

\end{document}